\documentclass[twocolumn,showpacs,floatfix]{revtex4}%
\usepackage{graphicx}%
\usepackage{amsmath}%
\usepackage{amsfonts}%
\usepackage{amssymb}
\usepackage{bm}

\def\d{{\partial}}

\def\e{{\epsilon}}

\def\w{{\omega}}
\def\a{{\alpha}}

\allowdisplaybreaks[4]

\begin{document}
\title{
Orbital Fluctuation Theory in Iron Pnictides: Effects of As-Fe-As Bond Angle,\\
Isotope Substitution, and $Z^2$-Orbital Pocket on the Superconductivity
}
\author{Tetsuro \textsc{Saito}$^{1}$, Seiichiro \textsc{Onari}$^{2}$,
and Hiroshi \textsc{Kontani}$^{1}$}

\date{\today }

\begin{abstract}
We study the pairing mechanism in iron pnictide superconductors
based on the five-orbital Hubbard-Holstein model.
Due to Fe-ion oscillations,
the $s$-wave superconducting (SC) state without 
sign reversal ($s_{++}$-wave state) is induced by orbital fluctuations
by using realistic model parameters.
The virtue of the present theory is that the famous empirical relation 
between $T_{\rm c}$ and the As-Fe-As bond angle is automatically explained,
since the electron-phonon ($e$-ph) coupling that creates the orbital 
fluctuations is the strongest when the As$_4$-tetrahedron is regular.
The negative iron isotope effect is also reproduced.
In addition, the magnitude of the SC gap on hole-pockets is predicted 
to be rather insensitive to the corresponding $d$-orbital ($xz/yz$- or 
$z^2$-orbital), which is consistent with the recent bulk-sensitive 
angle-resolved photoemission spectroscopy (ARPES) measurement for 
(Ba,K)Fe$_2$As$_2$ and BaFe$_2$(As,P)$_2$.
These obtained results indicate that 
the orbital-fluctuation-mediated $s_{++}$-wave state
is a plausible candidate for iron pnictides.
\end{abstract}

\address{
$^1$ Department of Physics, Nagoya University and JST, TRIP, 
Furo-cho, Nagoya 464-8602, Japan. 
\\
$^2$ Department of Applied Physics, Nagoya University and JST, TRIP, 
Furo-cho, Nagoya 464-8602, Japan. 
}
 
\pacs{74.70.Xa, 74.20.-z, 74.20.Rp}

\sloppy

\maketitle

\section{Introduction}

The understanding of the pairing mechanism in iron pnictide 
superconductors \cite{Kambara} has been a significant open problem.
By taking account of the Coulomb interaction at Fe-ions
and the nesting of the Fermi surfaces (FSs),
a fully-gapped sign-reversing $s$-wave state ($s_\pm$-wave state) 
has been proposed based on the spin fluctuation theories
 \cite{Kuroki,Mazin}.
Up to now, spin-fluctuation-mediated unconventional superconductivity
is believed to be realized in various metals,
such as high-$T_{\rm c}$ cuprates \cite{Moriya,Pines,Bickers},
$\kappa$-(BEDT-TTF)$_2$X \cite{Schmalian,Kino,Kondo},
and Ce$M$In$_5$ ($M$=Co,Rh,Ir) \cite{Takimoto-Ce115}.
To confirm the spin-fluctuation scenario in iron pnictides,
it is of significant importance to find evidences for the sign-reversal in the 
superconducting (SC) gap, 
and for the relationship between spin fluctuation strength and 
the SC transition temperature $T_{\rm c}$.

In principle,
spin-fluctuation-mediated superconductivity is fragile
against nonmagnetic impurities or randomness,
since the SC gap function has sign changes inevitably.
This is also true for iron pnictides, although the
FSs are disconnected and the SC gap is fully-gapped.
According to Ref. \cite{Onari-impurity}, 
decrease in $T_{\rm c}$ per $\rho_{\rm imp}=1\mu\Omega$cm reaches $\sim1$K
{\it independently of the impruity potential strength}.
Contrary to this expectation, 
the SC state is very robust against various impurities 
\cite{Sato-imp} and heavy-particle irradiations \cite{irradiation,Nakajima},
although carrier number dependence may exist \cite{FCZhang}.
Moreover, the spin-fluctuation-mediated superconductors are expected to 
show a ``resonance peak'' in the neutron inelastic scattering
as a reflection of sign-change in the SC gap,
as observed in high-$T_{\rm c}$ cuprates 
\cite{iikubo-sato,ito-sato,keimer-highTc}
and in Ce$M$In$_5$ ($M$=Co,Rh,Ir) \cite{res-115}.
However, the observed ``resonance-like'' peak structure in iron pnictides
\cite{christianson,qiu,keimer} 
is reproduced theoretically
by considering the strong correlation effect via quasiparticle 
damping, even in the conventional $s$-wave state 
without sign reversal ($s_{++}$-wave state)
 \cite{Onari-resonance}.

In BaFe$_2$(As$_{1-x}$P$_x$)$_2$,
$T_{\rm c}$ increases as $x$ decreases till the 
lattice structure transition occurs at $x=0.27$, and $T_{\rm c}$ is positively 
correlated to the spin-fluctuation strength for $x\ge0.33$ \cite{Ishida}.
On the other hand, $T_{\rm c}$ in LaFeAsO$_{1-x}$F$_x$ at $x=0.14$ 
increases from 26 K to 43 K by applying the pressure, whereas
spin-fluctuation strength observed by $1/T_1T$ measurement
is almost unchanged \cite{Fujiwara}.
Thus, the correlation between the spin-fluctuation strength 
and $T_{\rm c}$ seems to depend on compounds.

Considering these difficulties in the $s_\pm$-wave scenario,
we have proposed the orbital-fluctuation theory in Ref. \cite{Kontani},
by taking account of the $d$-orbital degree of freedom in iron pnictides.
It was found that large orbital fluctuations are induced by 
the electron-phonon ($e$-ph) interaction due to Fe-ion oscillations,
although they are not induced by Coulomb interaction alone.
Then, orbital-fluctuation-mediated $s_{++}$--wave SC state is realized 
\cite{Kontani,Yanagi}
even if $e$-ph interaction is smaller than that
estimated by the first principle study \cite{lambda-LDA}.
Existence of large ferro-orbital fluctuations is suggested
by prominent softening of shear modulus 
in Ba122 \cite{softening,Yoshizawa}.
Also, Raman spectroscopy \cite{Raman}, angle-resolved photoemission 
spectroscopy (ARPES) \cite{Dresden}, and optical conductivity
measurement \cite{optical}
highlight the importance of $e$-ph interaction.

In this paper, we analyze the five-orbital Hubbard-Holstein (HH) model
for iron pnictides in detail, by taking account of 
all the matrix elements of the $e$-ph interaction 
due to Fe-ion oscillations correctly.
It is found that a small $e$-ph interaction ($\lambda\lesssim0.15$) 
can induce substantial orbital fluctuations,
utilizing all five $d$-orbitals on the FSs efficiently.
Our main findings are as follows:
(i) Empirical relation between $T_{\rm c}$ and the As-Fe-As bond angle
(Lee plot) \cite{Lee} is automatically explained,
(ii) Experimental negative isotope effect 
\cite{Shirage1} is reproduced,
(iii) Magnitude of the SC gap on the $Z^2$-orbital
hole-pocket in (Ba,K)Fe$_2$As$_2$ and BaFe$_2$(As,P)$_2$
is comparable to the gap on other hole-pockets,
which is consistent with experiments \cite{Shimojima2}.
The range of model parameters for the $s_{++}$-wave SC state
becomes wider in the presence of small amount of nonmagnetic impurities
 \cite{Onari-impurity}.
In addition, the predicted orbital fluctuations had been observed 
as the softening of the elastic constants $C_{44}$ and $C_{E}$ 
in Ref. \cite{Yoshizawa}.

Here, we consider the ``orbital physics'' in iron pnictides:
It has been revealed that the ordered phase in mother compounds
is not a simple spin-density-wave (SDW) state,
but prominent orbital-density-wave coexists.
In fact, recent bulk-sensitive ARPES in BaFe$_2$As$_2$ below $T_{\rm N}$
had shown that the Fermi surface around $\Gamma$-point
are mainly composed of $xz$-orbital,
indicating the prominent nonequivalence between $xz$- and $yz$-orbitals
at the Fermi level in mother compounds
 \cite{Shimojima}.
Also, apparent in-plane anisotropy of resistivity had been observed
in detwinned Ba122 even above the structural transition temperature ($T_{\rm s}$),
suggesting the existence of nematic order as a pure electronic origin
 \cite{detwinned}.
These experimental facts indicate the existence of orbital fluctuations
even in (doped) superconducting compounds,
and therefore orbital-fluctuation-mediated $s_{++}$-wave SC state 
is expected to occur next to the orbital-density-wave state
in iron pnictides.


\section{Model and Hamiltonian}

We construct the five orbital HH model for iron pnictides,
by adding the electron-phonon (\textit{e}-ph) interaction to the 
Hubbard model in Ref.\cite{Kuroki}.
The Hubbard model is comprised of the potential term $\epsilon_{\mu}$,
hopping term $t^{\mu \nu}_{ij}$,
intraorbital Coulomb $U$, interorbital Coulomb $U'$,
Hund's coupling $J$, and pair hopping $J'$;
\cite{Kuroki}
\begin{eqnarray}
H_{\mathrm{Hub}} &= \displaystyle\sum _i \sum_{\mu}
\sum_{\sigma} \epsilon_{\mu} n_{i \mu \sigma} 
+ \sum_{ij} \sum_{\mu \nu} \sum_{\sigma} t^{\mu \nu}_{ij}c^{\dagger}_{i \mu \sigma} c_{j \nu \sigma} \notag \\
&+ \displaystyle\sum_i \biggl( U \sum_\mu n_{i \mu \uparrow} n_{i \mu \downarrow} + U' \sum_{\mu > \nu} \sum_{\sigma \sigma '} n_{i \mu \sigma} n_{i \nu \sigma '} \notag \\
&- J \displaystyle\sum_{\mu \neq \nu} \bm{S}_{i \mu} \cdot \bm{S}_{i \nu} 
+ J' \sum_{\mu \neq \nu} c^{\dagger}_{i \mu \uparrow} c^{\dagger}_{i \mu \downarrow} c_{i \nu \downarrow} c_{i \nu \uparrow} \biggl) , 
\label{eqn:Hamiltonian}
\end{eqnarray}
where $i$,$j$ denote the sites, and $\mu$,$\nu$ are the five $d$-orbitals.
We denote $Z^2$, $XZ$, $YZ$, $X^2-Y^2$, and $XY$ orbitals as 
1, 2, 3, 4, and 5 respectively:
The $X$- and $Y$-axes are parallel to the nearest Fe-As bonds,
and $Z$-axis is perpendicular to the FeAs plane \cite{Kuroki}.
The $XY$-coordinate is given by $-45^{\circ}$ rotation of 
the $xy$-coordinate spanned by the nearest Fe-Fe bonds around the $z$-axis.

\begin{figure}[!htb]
\includegraphics[width=0.9\linewidth]{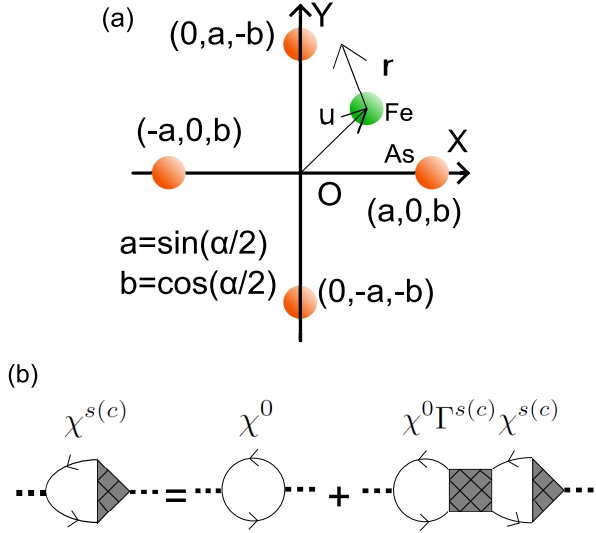}
\caption{
(Color online) 
(a) As$_4$ tetrahedron in iron pnictides shown along the $z$ axis.
Here, we put $R_{\text{Fe-As}}=1$.
The As-Fe-As bond angle $\alpha$ is illustrated in Fig. \ref{fig:A^2-alpha} (a).
(b) The diagrammatic expression for $\chi^{s(c)}$.}
\label{fig:fig1}
\end{figure}

Now, we derive the \textit{e}-ph interaction term 
due to the Einstein oscillation of Fe ions.
The Coulomb potential for the \textit{d} electron at $\bm{r}$ (with the origin at the center of the Fe ion)
due to the surrounding As$^{3-}$-ion tetrahedron is given by \cite{Kontani},
\begin{eqnarray}
\phi^{\pm} \left( \bm{r}; \bm{u} \right) = 3e^2 \displaystyle\sum^4_{s=1} 
\left\{ \left| \bm{r} + \bm{u} - \bm{R}_s^{\pm} \right|^{-1} 
-\left| \bm{r} - \bm{R}_s^{\pm} \right|^{-1} \right\} \notag \\
\approx \pm A \left[ 2 XZ \cdot u_X-2YZ \cdot u_Y + ( X^2 - Y^2)u_Z \right] ,
\label{eqn:V}
\end{eqnarray}
where $\bm{u}$ is the displacement vector of the Fe ion, $\bm{R}^{\pm}_s$ 
is the location of surrounding As ions in Fig. \ref{fig:fig1} (a);
When As$_4$ tetrahedron is regular tetrahedron ($\alpha=109.47^{\circ}$),
$a=\sqrt{2/3}$ and $b=\sqrt{1/3}$.
That is, $\bm{R}^{+}/R_{\text{Fe-As}} = ( \pm a , 0 , b )$ 
and $( 0 , \pm a , - b )$ for Fe$^{(1)}$,
and $\bm{R}^{-}/R_{\text{Fe-As}} = ( \pm a , 0 , - b )$ 
and $( 0 , \pm a , b )$ for Fe$^{(2)}$,
and $A=30e^2/ \sqrt{3} R^4_{\text{Fe-As}}$.
$R_{\text{Fe-As}}$ is the Fe-As bond length.
We neglect the As-ion oscillations since 
they do not induce substantial orbital fluctuations
unless very large $e$-ph interaction is assumed.

Nonzero matrix elements 
$\langle \mu | \phi | \nu \rangle = \sum_{\Xi}^{XYZ} v_{\mu \nu}^{\Xi} u_{\Xi}$ 
are given as
\begin{eqnarray}
v_{24}^X = v_{35}^X = \pm 2Aa^2/7, \notag \\
v_{34}^Y = - v_{25}^Y = \pm 2Aa^2/7, \notag \\
v_{22}^Z = - v_{33}^Z = \pm 2Aa^2/7, \notag \\
v_{12}^X = \pm 2Aa^2/7 \cdot (1/\sqrt{3}), \notag \\
v_{13}^Y = \mp 2Aa^2/7 \cdot (1/\sqrt{3}), \notag \\
v_{14}^Z = \mp 2Aa^2/7 \cdot (2/\sqrt{3}),
\label{eqn:Vmat}
\end{eqnarray}
where $a$ is the radius of \textit{d} orbital.
The obtained $e$-ph interaction does not couple to the charge density
since $v_{\mu \nu}^{\Xi}$ is trace-less.
Thus, the Thomas-Fermi screening for the coefficient $A$ is absent.
We stress that there are many nonzero (off-diagonal) elements 
in Eq. (\ref{eqn:Vmat})
due to the fact that As ions locate out of the Fe plane.
Then, the \textit{e}-ph interaction term is given by
\begin{equation}
H_{e-\mathrm{ph}} = \sum_{i} \sum_{\mu \nu} \sum_{\Xi} \sum_{\sigma} v_{\mu \nu}^{\Xi} c_{i \mu \sigma}^{\dagger} c_{i \nu \sigma} u_{i \Xi},
\end{equation}
%
which represents the orbital exchange process 
induced by the Fe ion displacement.
We will show later that substantial orbital fluctuations
involving all five orbitals are developed 
because of many nonzero elements in Eq. (\ref{eqn:Vmat}).

Next, we derive the phonon-mediated electron-electron interaction.
The local phonon Green function is 
\begin{equation}
D ( \omega_l ) = \frac{2 \bar{u}_0^2 \omega_D}{\omega_l^2 + \omega_D^2},
\end{equation}
which is the Fourier transformation of 
$ \langle T_{\tau} u_{\mu} ( \tau ) u_{\mu} ( 0 ) \rangle $ $( \mu = X,Y,Z )$.
$\omega_D$ is the phonon frequency, and 
$\bar{u}_0 = \sqrt{ \hbar / 2M_{\text{Fe}} \omega_D}$ is the 
uncertainty in position for Fe ions; $\bar{u}_0 = 0.044$ 
${\buildrel _{\circ} \over {\mathrm{A}}}$ for $\w_{\rm D}=0.02$ eV.
$\omega_l = 2 \pi lT$ is the boson Matsubara frequency.

For both Fe$^{(1)}$ and Fe$^{(2)}$, the phonon mediated interaction is obtained as
\begin{equation}
H^{\mathrm{ph}}_{\text{e-e}}= \sum_{\mu \nu \mu ' \nu '} \sum_{i} \sum_{\sigma \sigma '} V_{\mu \nu,\mu ' \nu '} (\omega_l)
c_{i \mu \sigma}^{\dagger} c_{i \nu \sigma} c_{i \mu ' \sigma '}^{\dagger} c_{i \nu ' \sigma '}.
\label{eqn:He-e}
\end{equation}
In Eq. \eqref{eqn:He-e}, nonzero $V_{\mu \nu,\mu ' \nu '}$ are given as
\begin{eqnarray}
V_{24,24} = V_{34,34} = V_{22,22} = V_{33,33} = -V_{22,33} = - g (\w_l ), \notag \\
V_{25,25} = V_{35,35} = V_{24,35} = -V_{25,34} = - g (\w_l ), \notag \\
V_{12,35} = V_{13,25} = V_{12,24} = - V_{13,34} = -(1/\sqrt{3}) g (\w_l ), \notag \\
V_{12,12} = V_{13,13} = -(1/3) g (\w_l ), \notag \\
V_{14,33} = -V_{14,22} = -(2/\sqrt{3}) g (\w_l ), \notag \\
V_{14,14} = -(4/3) g (\w_l ),
\label{eqn:effpot}
\end{eqnarray}
where $g ( \w_l ) \equiv (2Aa^2/7)^2 D ( \w_l )$.
The following relations hold;
$V_{lm,l'm'}=V_{ml,l'm'}=V_{lm,m'l'}$ and $V_{lm,l'm'}=V_{l'm',lm}$.
In our earlier investigation \cite{Kontani}, 
we have estimated $g(0) \approx 0.4$eV
for $R_{\mathrm{Fe-As}} \approx 2.4$ ${\buildrel _{\circ} \over {\mathrm{A}}}$,
$a \approx 0.77$ ${\buildrel _{\circ} \over {\mathrm{A}}}$, and $\omega_D \approx 0.018$eV.
We also obtain $g(0)\approx 0.33$eV for $\omega_D \approx 0.02$eV.

In our earlier calculation \cite{Kontani}, 
only the first line of Eq. \eqref{eqn:effpot} was taken into account
since the weight of orbitals 1 and 5 on the FSs is small.
In this paper,
however, we will show later that orbital fluctuations increase 
substantially if all the interactions in Eq. \eqref{eqn:effpot} 
are taken into account correctly.
This is one of the main message in the present work.

Now, we perform the RPA for $H_{\rm Hub}+H_{\text{e-e}}^{\rm ph}$.
The irreducible susceptibility in the five orbital model is given by
\begin{equation} 
\chi^0_{ll',mm'} \left( q \right) = - \frac{T}{N} \sum_k G_{lm}^0 \left( k+ q \right) G_{m' l'}^0 \left( k \right),
\end{equation}
where $\hat{G}^0 ( k ) = [ i \epsilon_n + \mu - \hat{H}^0_{\bm{k}} ]^{-1}$
is the \textit{d} electron Green function in the orbital basis,
$q = ( \bm{q}, \omega_l )$, $k=( \bm{k} , \epsilon_n)$,
and $\epsilon_n = (2n + 1) \pi T$ is the fermion Matsubara frequency.
$\mu$ is the chemical potential, and $\hat{H}^0_{\bm{k}}$ is the kinetic term 
of Eq. \eqref{eqn:Hamiltonian}.
Then, the susceptibilities for spin and charge sectors in the RPA are given by \cite{Takimoto}
\begin{gather} 
\hat{\chi}^{s} \left( q \right) = \frac{\hat{\chi}^0 \left( q \right)}{1 - \hat{\Gamma}^{s} \hat{\chi}^0 \left( q \right)}, \\
\hat{\chi}^{c} \left( q \right) = \frac{\hat{\chi}^0 \left( q \right)}{1 - \hat{\Gamma}^{c} (\omega_l) \hat{\chi}^0 \left( q \right)},
\end{gather}
where
\begin{equation}
\Gamma_{l_{1}l_{2},l_{3}l_{4}}^s = \begin{cases}
U, & l_1=l_2=l_3=l_4 \\
U' , & l_1=l_3 \neq l_2=l_4 \\
J, & l_1=l_2 \neq l_3=l_4 \\
J' , & l_1=l_4 \neq l_2=l_3
\end{cases}
\end{equation}
\begin{equation}
\hat{\Gamma}^c ( \omega_l )= -\hat{C} - 2\hat{V}( \w_l ),
\end{equation}
\begin{equation}
C_{l_{1}l_{2},l_{3}l_{4}} = \begin{cases}
U, & l_1=l_2=l_3=l_4 \\
-U'+2J , & l_1=l_3 \neq l_2=l_4 \\
2U' - J, & l_1=l_2 \neq l_3=l_4 \\
J' , & l_1=l_4 \neq l_2=l_3
\end{cases}
\end{equation}
Here, we neglect the ladder-diagram for phonon-mediated interaction
because of the relation $\omega_D \ll W_{\mathrm{band}}$\cite{Kontani,Yanagi}.
The Bethe-Salpeter equation for $\chi^{s(c)}$ is given in Fig. \ref{fig:fig1} (b).

\begin{figure}[!htb]
\includegraphics[width=0.9\linewidth]{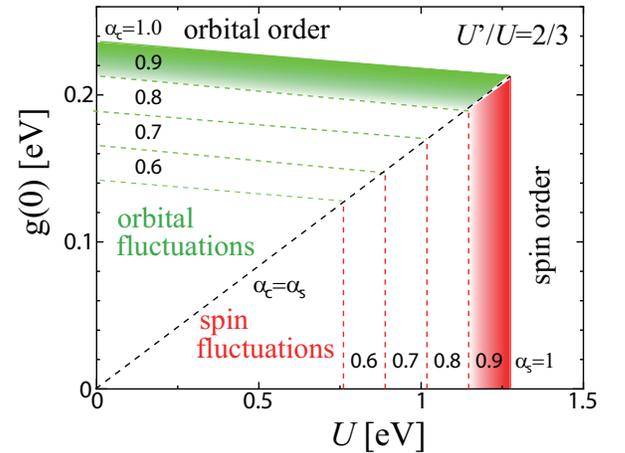}
\caption{
(Color online) 
Obtained $U$-$g(0)$ phase diagram for $n=6.1$.
Near the orbital-density-wave boundary, $s_{++}$-wave SC state
is realized by orbital fluctuations.
}
\label{fig:U-g0}
\end{figure}

Hereafter, we assume that $J=J'$ and $U=U'+2J$,
and fix the ratio $J/U=1/6$.
Figure \ref{fig:U-g0} shows the $U$-$g(0)$ phase diagram for $n=6.1$
given by the RPA.
$\alpha_{s(c)}$ is the spin (charge) Stoner factor, which is given by the maximum eigenvalue of
$\hat{\Gamma}^{s(c)} \hat{\chi}^0 ( \bm{q},0 )$.
The transition line for the spin (orbital) order is given by the condition $\alpha_{s(c)}=1$.
Note that this phase diagram is independent of $\omega_D$
since $\alpha_{s(c)}$ is free from $\omega_D$.
For $U=1$ eV, the critical value $g_{\mathrm{cr}}(0)$ for $\alpha_c = 1$ is 0.22,
which means that dimensionless coupling constant is $\lambda_{\mathrm{cr}} \equiv g_{\mathrm{cr}}(0) N(0) \sim 0.15$, where $N(0)$ is the density of states 
per spin at the Fermi level.
Hereafter, the unit of energy is eV.

In our earlier work \cite{Kontani},
we have shown that $g_{\mathrm{cr}}(0) \sim 0.4$
when only the first line of Eq. \eqref{eqn:effpot} is considered.
However, we stress that $g_{\mathrm{cr}} (0)$ in Fig. \ref{fig:U-g0} in the present paper is almost halved.
This result means that strong orbital fluctuations are induced by much
smaller $g(0)$ by utilizing all five $d$-orbitals on the FSs efficiently.
Therefore, we take all the interaction in Eq. \eqref{eqn:effpot} into account in later calculations.

%
\begin{figure}[!htb]

\includegraphics[width=0.65\linewidth]{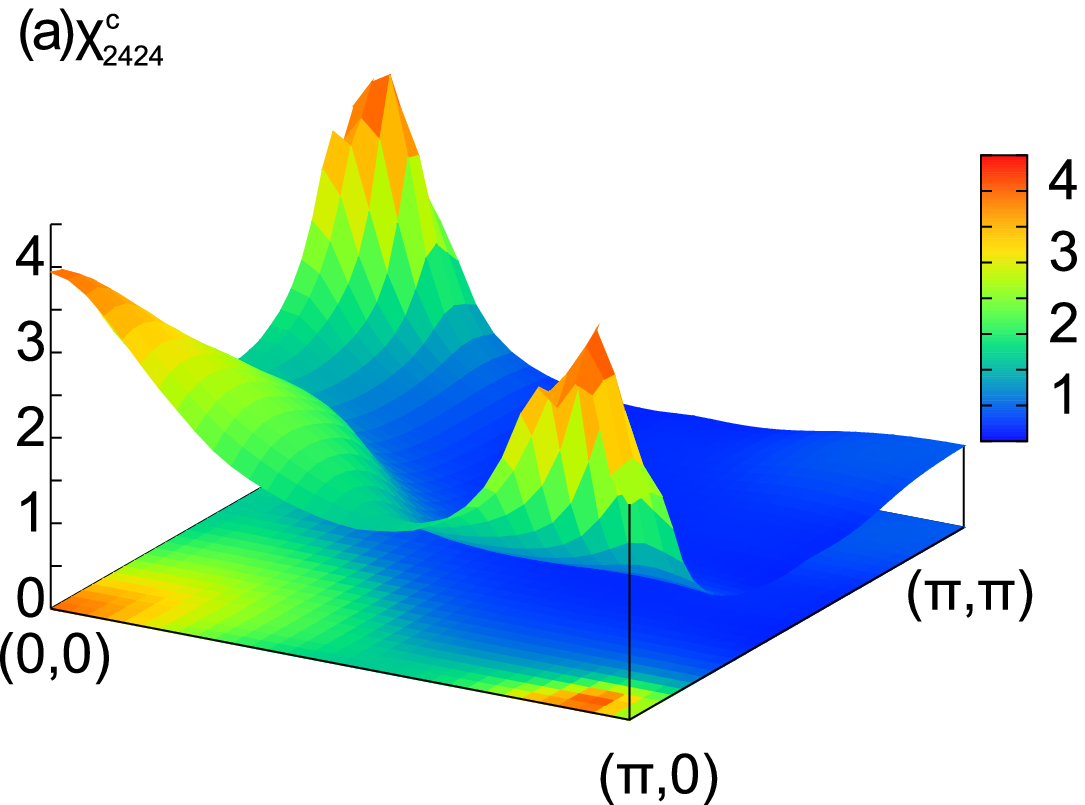}
\includegraphics[width=0.65\linewidth]{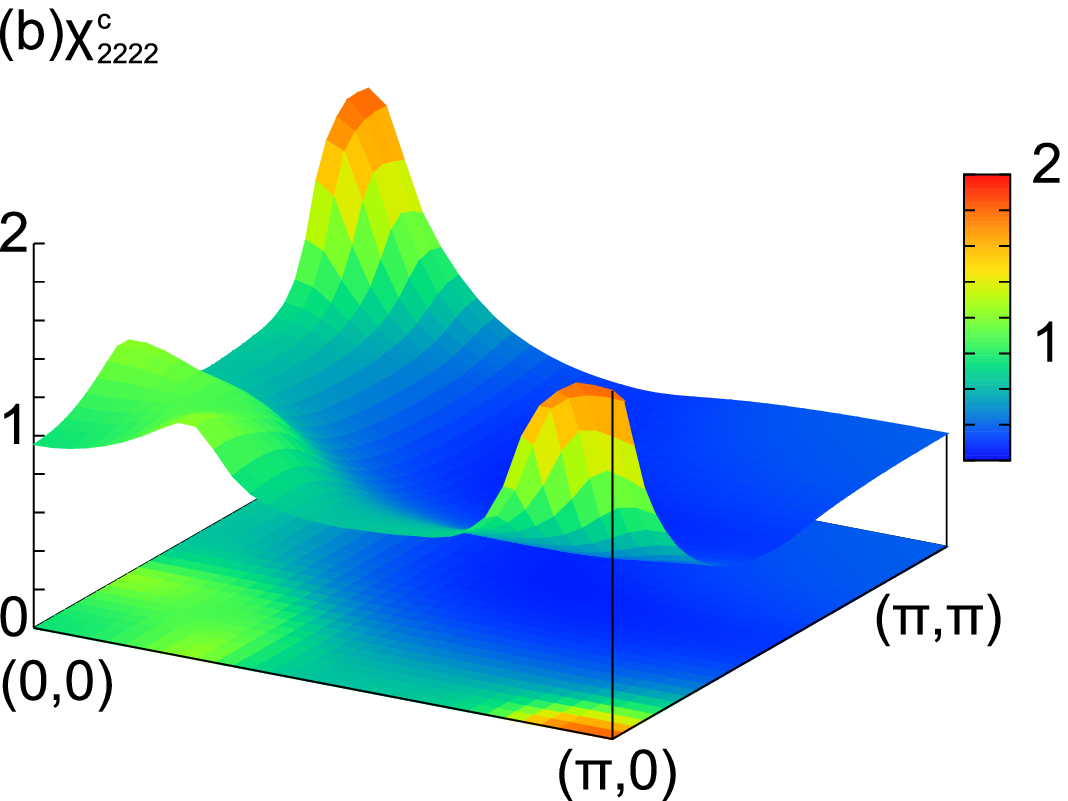}
\includegraphics[width=0.65\linewidth]{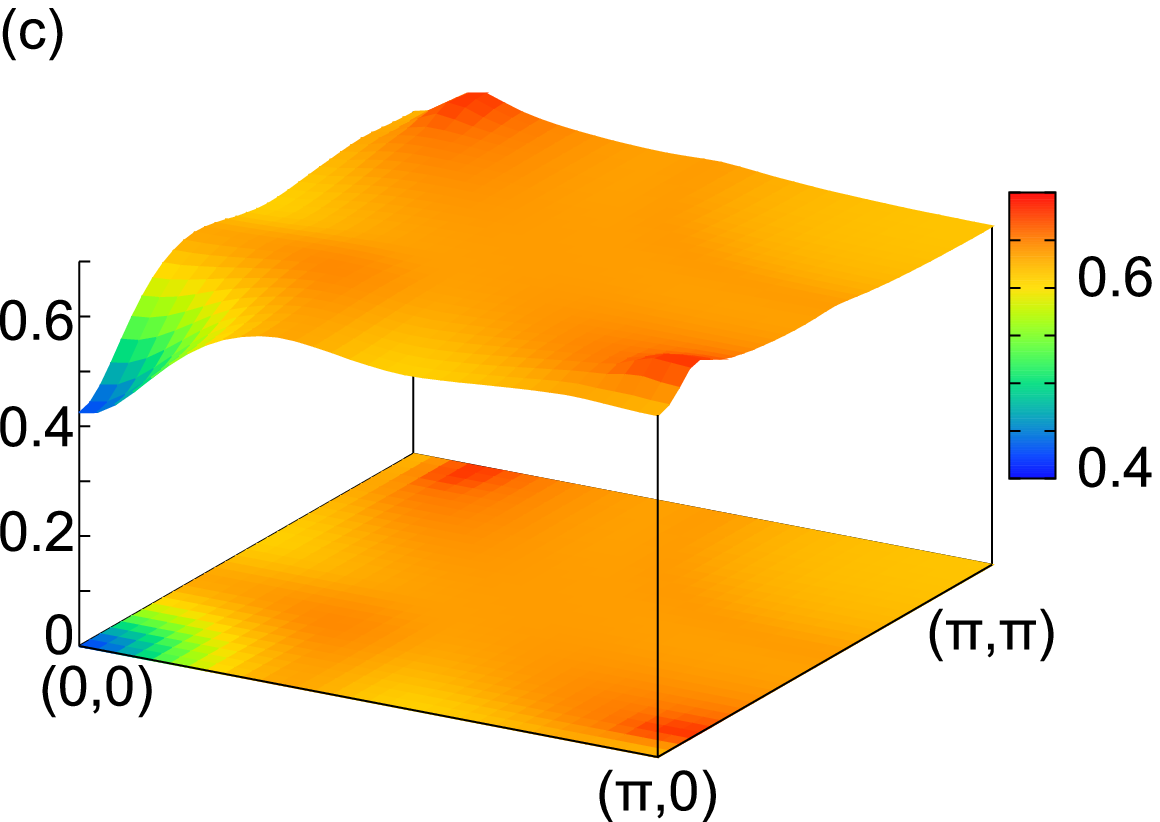}
\caption{
(Color online) 
Obtained (a) $\chi^c_{24,24}( \bm{q},0)$ and (b) $\chi^c_{22,22}( \bm{q},0)$,
for $n=6.1$, $U=1$, $T=0.02$, and $g(0)=0.21$.
Note that $\chi^c_{24,24}( \bm{q} , 0)=\chi^c_{34,34}( \bm{q}' , 0)$
and $\chi^c_{22,22}( \bm{q} , 0)=\chi^c_{33,33}( \bm{q}' , 0)$,
where $\bm{q}'$ is given by the rotation of $\bm{q}$ by $\pi/2$.
In (c), charge susceptibility $\chi^c(\bm{q},0)
\equiv \sum_{l,m}\chi^c_{ll,mm}(\bm{q},0)$ is shown.
We use 2048 Matsubara frequencies.}
\label{fig:chi_c}
\end{figure}

Figure \ref{fig:chi_c} shows the obtained $\chi^c_{ll',mm'}( \bm{q},0)$
for $(ll',mm') = (24,24)$ and $(22,22)$, respectively.
Used parameters are $n=6.1$, $U=1$, $T=0.02$,
and $g(0)=0.21$, which correspond to $\alpha_c=0.98$.
$\chi^c_{24,24}$ ($\chi^c_{34,34}$) and $\chi^c_{22,22}$ ($\chi^c_{33,33}$)
are the most divergent channels for electron doped case.
First, we discuss $\chi^c_{24,24}$ in Fig. \ref{fig:chi_c} (a):
It has the largest peak near $\bm{Q} = ( \pi , 0 )$,
which comes from the nesting between FS3,4 and FS1,2
in Fig. \ref{fig:gap} (c), and the multiple scattering by $V_{24,24}$.
Its second largest peak near $\bm{q} = (0,0)$
originates from the forward scattering by $V_{24,24}$
in the FS3 or FS4 that is composed of 2-4 orbitals.
$\chi^c_{24,24} ( \bm{0},0 )$ and $\chi^c_{24,24} ( \bm{Q} ,0)$ are comparable,
meaning that the ferro-orbital and antiferro-orbital orders are
highly frustrated. 
Note that $\chi^c_{24,24}(\bm{Q},0) \ll \chi^c_{24,24} (\bm{0},0)$ 
if we consider only the first line of Eq. \eqref{eqn:effpot} 
as demonstrated in Ref.\cite{Kontani}.

We also discuss $\chi^c_{22,22}$ in Fig. \ref{fig:chi_c} (b):
It has the largest peak near $\bm{Q} = ( \pi ,0)$,
which arises due to nesting between FS3,4 and FS1,2.
$\chi^c_{22,22}$ also has a slightly lower peak around $(0,0)$
that originates from the forward scattering in each FS.
We note that the large enhancement in $\chi^c_{24,24}$ and $\chi^c_{34,34}$ 
($\chi^c_{22,22}$ and $\chi^c_{33,33}$)
is caused by in-plane (out-of-plane) Fe-ion oscillations.
On the other hand, the total charge susceptibility 
$\chi^c(\bm{q},0)\equiv \sum_{l,m}\chi^c_{ll,mm}(\bm{q},0)$ is not enhanced 
as shown in Fig. \ref{fig:chi_c} (c), because of the relation 
$\chi^c_{22,22}(\bm{q},0)\approx-\chi^c_{22,33}(\bm{q},0)$ \cite{Kontani}.
Therefore, the origin of the superconductivity in the present model is 
not charge fluctuations, but orbital fluctuations
that can develop without cost of the Coulomb potential energy.

Finally, we discuss the softening in the elastic constants
due to orbital fluctuations.
Recently, Yoshizawa {\it et al.} have observed large softening in
$C_{44}$, $C_{66}$, and $C_{\rm E}$ in Ba(Fe,Co)$_2$As$_2$ \cite{Yoshizawa}.
The corresponding strains for $C_{44}$, $C_{66}$, and $C_{\rm E}$
are $\e_{XZ}$, $\e_{XY}$, and $\e_{XX}-\e_{YY}$, respectively.
Using the point-charge model, one can verify that
the strains $\e_{\mu\nu}$ ($\mu,\nu=X,Y,Z$) induce the quadrupole 
potential on each Fe ion; $\phi_{\mu\nu}\propto \mu\nu$.
The corresponding matrix elements 
$ \langle l|\phi_{XZ}|m\rangle\equiv \phi^{XZ}_{lm}$ and
$ \langle l|\phi_{XX}-\phi_{YY}|m\rangle \equiv \phi^{X^2-Y^2}_{lm}$
are proportional to $|v_{lm}^X|$ and $|v_{lm}^Z|$, respectively,
where $v_{lm}^\Xi$ are described in Eq. (\ref{eqn:Vmat}).
In the linear response theory, 
the enhancement in $C_{44}^{-1}$ ($C_{\rm E}^{-1}$) is proportional to
the quadrupole susuceptibility
$\sum_{ll'mm'}\phi_{ll'}^{\Xi}\chi^c_{ll',mm'}(\bm{0},0)\phi_{mm'}^{\Xi}$ with 
$\Xi=XZ$ ($\Xi=X^2-Y^2$).
As shown in Fig. \ref{fig:chi_c}, both $\chi^c_{24,24}$ and 
$\chi^c_{22,22}$ evolve in the present study.
Considering that $\phi_{lm}^{XZ}\propto |v_{lm}^{X}|$ is finite for $l,m=2,4$,
we find that the softening in $C_{44}$ is induced by
fluctuations in the $(2,4)$-channel.
In the same way, the softening in $C_{\rm E}$ is induced by
fluctuations in $(2,2)$-, $(3,3)$-, and $(2,3)$-channels.
These results are consistent with the reports in Ref. \cite{Yoshizawa}.
That is, theoretically predicted orbital fluctuations 
in Figs. \ref{fig:chi_c} (a) and (b)
had been confirmed experimentally.


\section{Eliashberg gap equation
\label{sec:Eliash}}

In this section, we analyze the following linearized Eliashberg equation 
using the RPA
by taking account of both the spin and orbital fluctuations 
on the equal footing \cite{Takimoto}:
\begin{equation}
 \begin{split}
\lambda_E \Delta_{ll'} (k) = \frac{T}{N} \sum_{k',m_i} W_{l m_1,m_4l'}(k-k') G_{m_1m_2}^0(k') \\
\times \Delta_{m_2m_3}(k') G_{m_4m_3}^0(-k'),
 \end{split}
\end{equation}
where,
\begin{equation}
\hat{W}(q) = - \frac{3}{2} \hat{\Gamma^s} \hat{\chi^s} \hat{\Gamma^s} 
+ \frac{1}{2} \hat{\Gamma^c} \hat{\chi^c} \hat{\Gamma^c}
- \frac{1}{2} (\hat{\Gamma^s} - \hat{\Gamma^c} ),
\end{equation}
for the singlet states.
$\lambda_E$ is the eigenvalue of the gap equation,
which approaches unity as $T\rightarrow T_c$.
Hereafter, we use 64$^2$ $\bm{k}$ meshes, 
and 1024 or 2048 Matsubara frequencies.
We perform the calculation at relativity high temperatures ($T \ge 0.02$)
since the number of meshes is not enough for $T<0.02$.
%
\begin{figure}[!htb]
\includegraphics[width=0.49\linewidth]{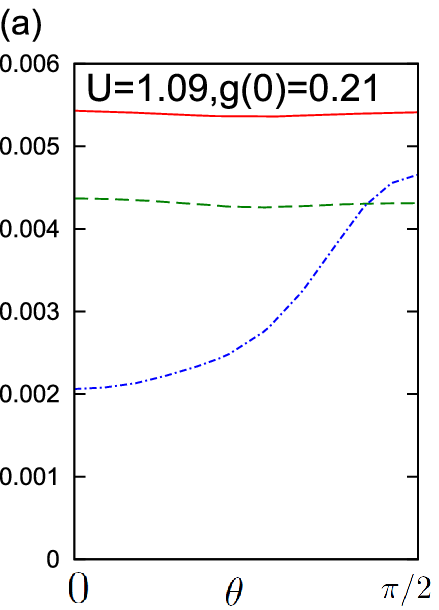}
\includegraphics[width=0.49\linewidth]{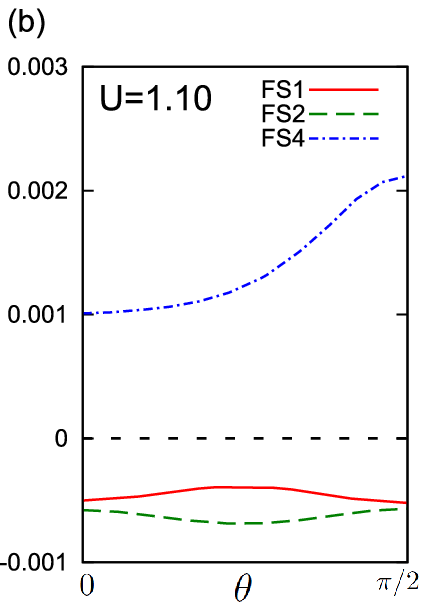}
\includegraphics[width=0.49\linewidth]{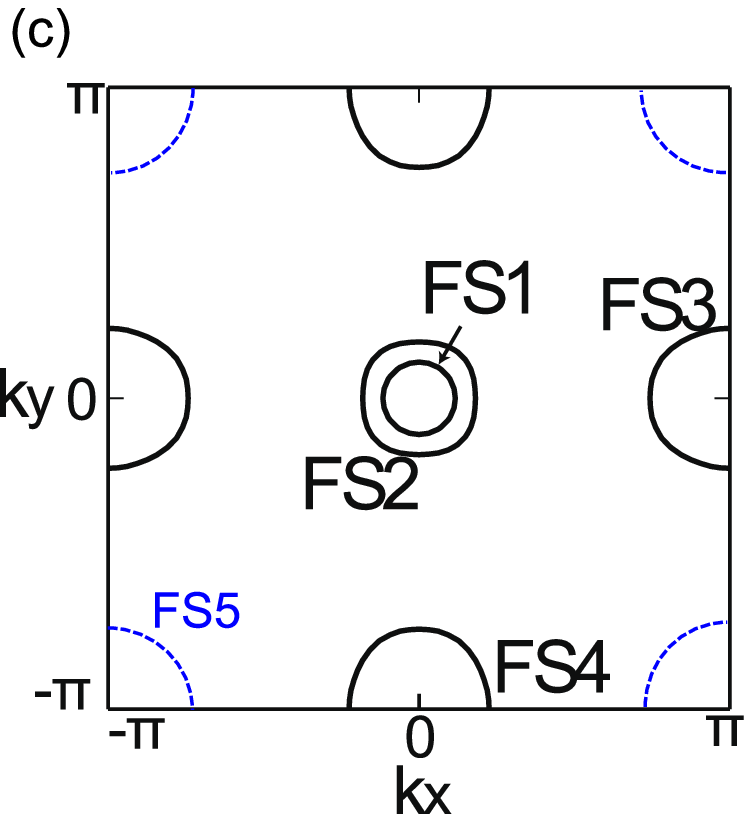}

\caption{
(Color online) 
(a) (b) Obtained SC gap functions for (a)$U=1.09$ and (b)$U=1.10$, 
respectively.
We put $g(0) = 0.21$ ($\alpha_c = 0.98$),
$T=0.02$, and $\omega_D=0.02$.
They are normalized as $N^{-1} \sum_{\bm{k},lm} | \Delta_{lm} ( \bm{k} ) |^2 = 1$.
We use 2048 Matsubara frequencies.
(c) FSs in the unfolded Brillouin zone.
FS1,2 (FS3,4) are composed of $2,3$-orbitals ($2,3,4$-orbitals).
FS5 is composed of $1$-orbital in (Ba,K)Fe$_2$As$_2$ and BaFe$_2$(As,P)$_2$,
which will be discussed in Sec. \ref{sec:z2}.
Note that FS5 moves to $(0,0)$ in the folded zone.
}
\label{fig:gap}
\end{figure}

Figure \ref{fig:gap} (a) and (b) show the SC gap on the FSs in the band representation
for $U=1.09$ and $U=1.10$, respectively.
We put $n=6.1$, $T=0.02$, $\omega_D=0.02$, and
$g(0) = 0.21$, which correspond to $\alpha_c = 0.98$.
The horizontal axis is the azimuth angle for $\bm{k}$ point 
with the origin at $\mathrm{\Gamma}$(M) point for FS1,2 (FS4).
For $U = 1.09$, the $s_{++}$ state is realized by orbital fluctuations \cite{Kontani}.
On the other hand, the $s_{\pm}$ state is realized for $U = 1.10$
since the spin fluctuations dominate the orbital fluctuations.
In this case, the boundary of the $s_{++} \to s_{\pm}$ phase transition 
is uniquely defined since the obtained gap functions are always full-gap.
If one introduces low concentration of impurities, 
the $s_{++}$-wave state is realized even for $U>1.1$ \cite{Kontani},
and moreover, the transition becomes full-gap 
$s_{++} \to $nodal s $\to$ full-gap $s_{\pm}$ \cite{Onari-future}.
For a quantitative study of the line nodes 
observed in several 122 compounds \cite{Matsuda},
the 3-dimensionality of the FSs may be indispensable.

%
\begin{figure}[!htb]
\includegraphics[width=0.90\linewidth]{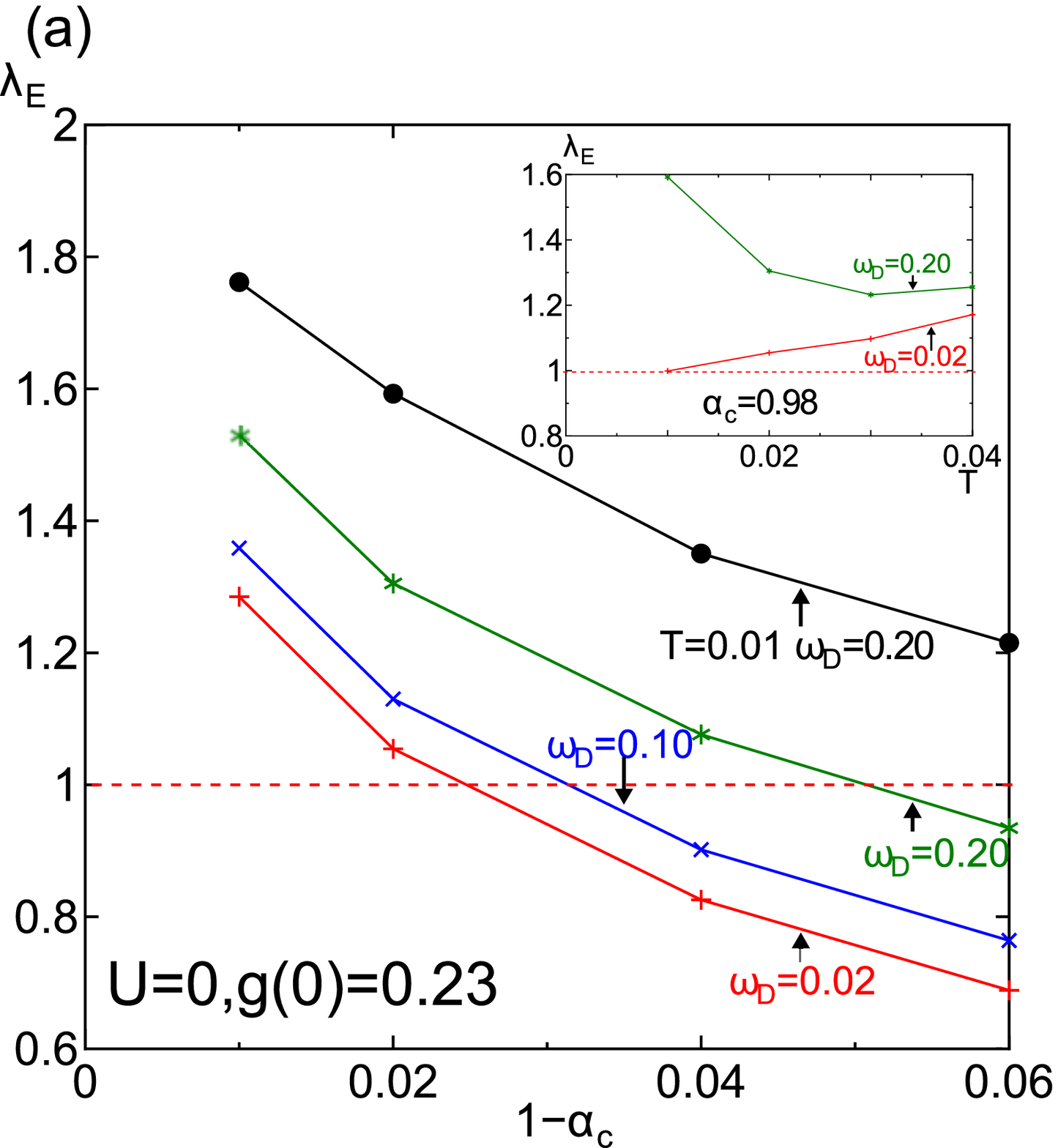}
\includegraphics[width=0.90\linewidth]{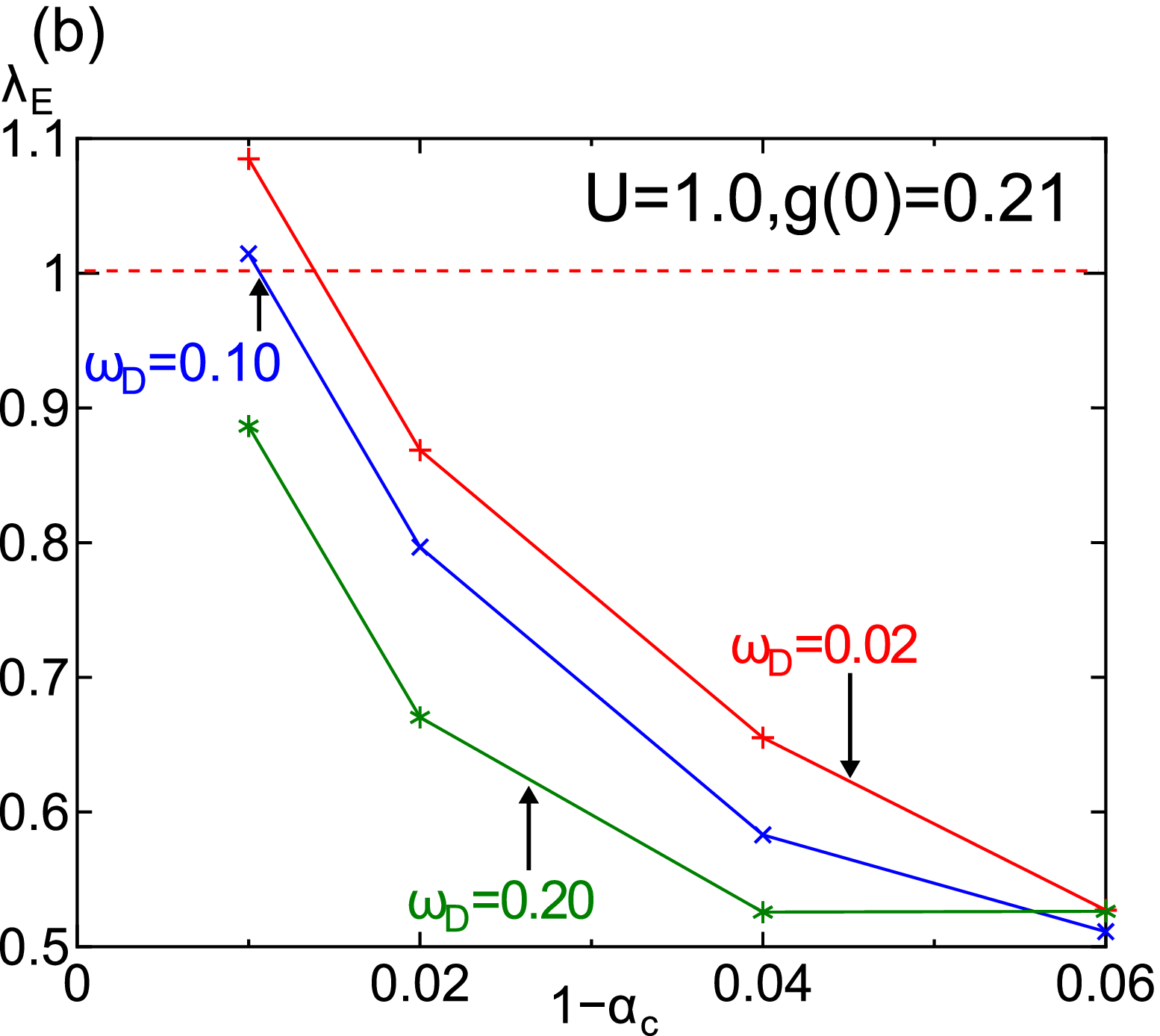}

\caption{
(Color online) 
Obtained $\lambda_E$ as function of $1-\alpha_c$ 
for (a)$U=0$ and (b)$U=1$, respectively, at $T=0.02$.
The former (latter) corresponds to the positive (negative) isotope effect.
We use 2048 Matsubara frequencies.
Inset in (a): $T$-dependence of $\lambda_E$ for $\alpha_c = 0.98$.
}
\label{fig:ac-lambda}
\end{figure}

Figure \ref{fig:ac-lambda} shows the
obtained $\lambda_E$ for (a) $U=0$ and (b) $U=1$, respectively, at $T=0.02$.
In case (a), $\lambda_E$ exceeds unity for $\omega_D=0.02$ 
when $1- \alpha_c \simeq 0.02$.
$\lambda_E$ increases as $\omega_D$ increases.
Considering the relation $\omega_D \propto 1/ \sqrt{M}$, 
this result means the positive isotope effect.
In case (b), $\lambda_E$ is larger than unity for $\omega_D = 0.02$ if $1 - \alpha_c \simeq 0.01$.
We stress that $\lambda_E$ decreases as $\omega_D$ increases,
which means that the negative isotope effect is realized.

%
Now, we discuss the origin of the negative isotope effect in case (b).
In the BCS theory, $T_c$ in a single band model is given by \cite{Morel}
\begin{equation}
T_c = 1.13 \omega_D \exp{\left( - \frac{1}{\lambda - \mu^*} \right)},
\label{eqn:Tc-isotope}
\end{equation}
where $\lambda = gN(0)$ is the dimensionless coupling constant, 
and $\mu^{*}$ is given as
\begin{equation}
\mu^* = \frac{\mu}{1 + \mu \ln{ \left( \bar{\epsilon} / \omega_D \right)}},
\end{equation}
where $\mu = UN(0)$.
$\mu^*$ is called the Morel-Anderson pseudo-potential.
In general, $\mu^{*} \ll \mu$ since the limit of Coulomb interaction $\bar{\epsilon}$
is much larger than $\omega_D$.
Now, we derive the coefficient $\beta =-\d \ln{T_c} / \d \ln M$
($T_{\rm c}\propto M^{-\beta}$), where $M$ is mass of Fe ion.
By differentiating Eq. \eqref{eqn:Tc-isotope} by $M$ using the relation $\omega_D \propto 1/ \sqrt{M}$,
we obtain
\begin{equation}
\beta = \frac{1}{2} \left[1- \frac{\mu^{*2}}{\left( \lambda - \mu^* \right)^2} \right] .
\label{eqn:beta}
\end{equation}
The value of $\beta$ decreases from $1/2$ as $\mu^*$ increases,
and becomes negative when $\mu^*$ is larger than $\lambda/2$.
This can be realized when $U$ is relatively large.
Therefore, the negative isotope effect ($\beta <0$) 
in Fig. \ref{fig:ac-lambda} (b) is caused by the reduction in $\mu^*$.
In other words, negative isotope effect originates from the enhancement of the retardation effect.

Note that Eq. (\ref{eqn:beta}) is valid only for one-band model.
In order to obtain the correct coefficient $\beta$ in iron pnictides,
we have to analyze the five-orbital model.

%
In case of Fig. \ref{fig:ac-lambda} (b), $\lambda_E$ exceeds unity only when $\alpha_c \sim 0.99$.
However, $\lambda_E$ can reach unity for smaller $\alpha_c$
when the temperature is much lower than $\omega_D = 0.02$.
To explain this behavior, we analyze the following single band gap equation
\cite{Allen}:
\begin{equation}
\lambda_E \Delta = T \sum_{\bm{k}, n} \frac{\Delta}{\epsilon_{n}^2 + \epsilon_{\bm{k}}^2} V (i \epsilon_{n}).
\end{equation}
Here we consider the BCS approximation $V(k, i \e_n ) = -g \theta ( \omega_D - | \e_n - \pi T | )$.
After carrying out the $\bm{k}$ summation, we obtain
\begin{equation}
\lambda_E \Delta = 2gN (0) \sum_{l=0}^{n_c} \frac{1}{2l+1} \Delta,
\label{eqn:18}
\end{equation}
where $(2n_c+1) \pi T = \omega_D$.
When $\omega_D \gg T$ (i.e. $n_c \gg 1$), Eq. \eqref{eqn:18} is solved as 
\cite{Allen}:
\begin{equation}
\lambda_E = gN (0) \ln (1.13 \omega_D/T ) .
\label{lowT-lambda}
\end{equation}
Therefore, $\lambda_E$ diverges logarithmically at low temperatures.
On the other hand, when $\omega_D \simeq T$, Eq. \eqref{eqn:18} is solved by putting $n_c=0$ as
\begin{equation}
\lambda_E = 2gN (0) .
\label{highT-lambda}
\end{equation}
In this case, $\lambda_E$ does not depend on $T$.
Inset of Fig. \ref{fig:ac-lambda} (a) shows the $T$-dependence of 
$\lambda_E$ for $U=0$ and $g(0) = 0.23$ ($\alpha_c = 0.98$).
For $\omega_D = 0.20$, $\lambda_E$ increases at low temperatures,
in accordance with Eq. \eqref{lowT-lambda}.
In contrast, $\lambda_E$ slightly decreases for $\omega_D = 0.02$
at low temperature, which might be due to the smaller size of 
$\bm{k}$- or $\w$-meshes.
Unfortunately, we can not perform the calculation below 100K
since the numbers of $\bm{k}$- and $\w$-meshes are not sufficient.
However, even if $\omega_D = 0.02$, $\lambda_E$ will increase below 
$T \sim 20$K.
Therefore, $\lambda_E$ is expected to exceed unity at low temperatures 
even if $\alpha_c \ll 0.99$.

Finally, we briefly discuss the effect of self-energy $\Sigma$,
which has been dropped in the present study.
The quasiparticle damping $\gamma$ (= imaginary part of $\Sigma$)
reduces both $\alpha_{s(c)}$ and $\lambda_{\rm E}$.
However, the dimensionless coupling constant for $\alpha_c=0.98$
in the FLEX approximation is only $\lambda\equiv g(0)N(0)\sim0.2$
\cite{Onari-future}.
Thus, the present orbital-fluctuation scenario is justified
even if the self-energy correction is taken into account.
In the FLEX approximation,
$g_{\rm cr}(0)$ for orbital-density-wave state and $U_{\rm cr}$ for SDW state
become infinity in 2-dimensional systems 
since Mermin-Wagner theorem is satisfied
 \cite{Kontani-imp}.


\section{Crystal Structure and T$_\text{c}$
\label{sec:alpha}}

\begin{figure}[!htb]
\includegraphics[width=0.39\linewidth]{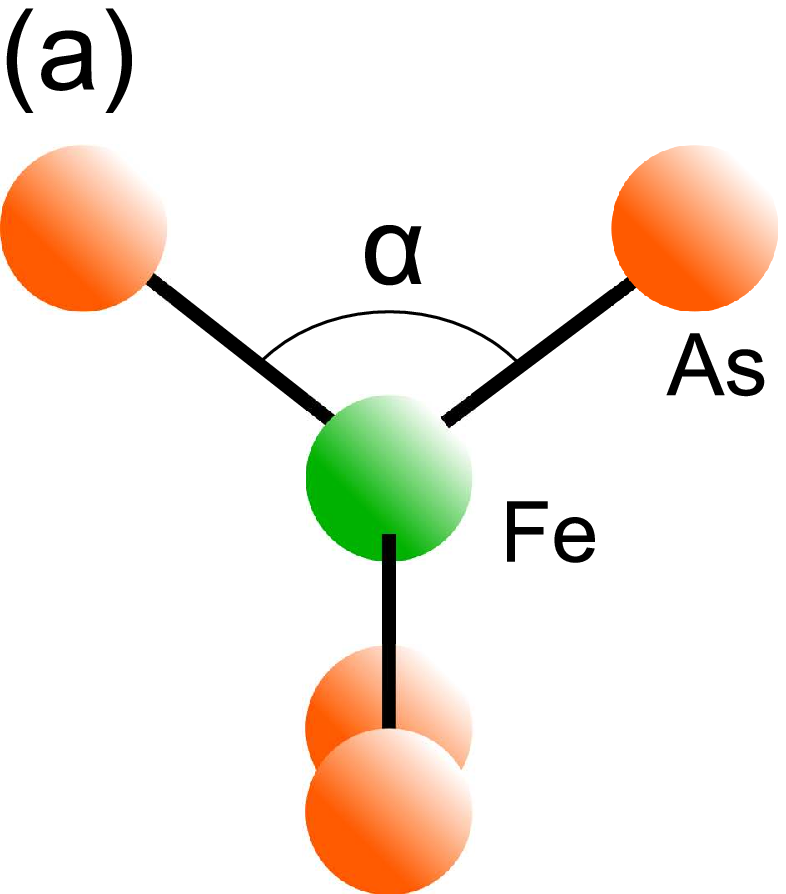}
\includegraphics[width=0.59\linewidth]{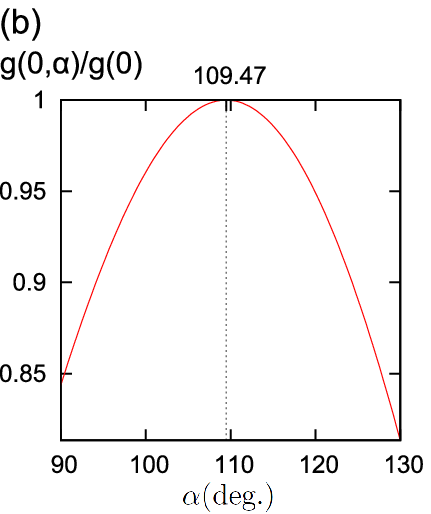}

\caption{
(Color online) 
(a) The definitions of the As-Fe-As bond angle $\alpha$.
(b) $g(0, \alpha )/g(0)$ as a function of $\alpha$.}
\label{fig:A^2-alpha}
\end{figure}

In previous sections, we assumed that the As$_4$ tetrahedron forms a regular tetrahedron.
However, it is well-known that As-Fe-As bond angle $\alpha$,
which is shown in Fig. \ref{fig:A^2-alpha}(a),
closely relates on $T_c$ experimentally \cite{Lee}.
Here, we extend the theory for general bond angle $\alpha$.
First, we derive the \textit{e}-ph interaction for general $\alpha$.
In this case, the potential $\phi ( \bm{r}; \bm{u})$ 
in Eq. \eqref{eqn:V} is not changed except for $A$;
$A$ is changed as
\begin{equation}
A= \frac{30e^2}{\sqrt{3}R^4_{\text{Fe-As}}} \left( \frac{3 \sqrt{3}}{2} \sin^2{\frac{\alpha}{2}} \cos{\frac{\alpha}{2}} \right).
\label{eq:A(alpha)}
\end{equation}
When the As$_4$ tetrahedron forms the regular tetrahedron
($\alpha = \alpha_0 \equiv 109.47^{\circ}$),
the term in the brackets in Eq. \eqref{eq:A(alpha)} takes the maximum value; unity.
Thus, the effective interaction for $\alpha$ is given by
$g(0, \alpha ) = g(0) [(3 \sqrt{3}/2) \sin^2(\alpha /2) \cos(\alpha/2)]^2$,
where $g(0) \equiv g (0, \alpha_0)$
In Fig. \ref{fig:A^2-alpha}(b), we show $g(0,\alpha )/ g(0)$ as a function of $\alpha$.
When $\alpha$ deviates from $\alpha_0$, $A^2$ decreases rapidly.
%
%

\begin{figure}[!htb]
\includegraphics[width=0.65\linewidth]{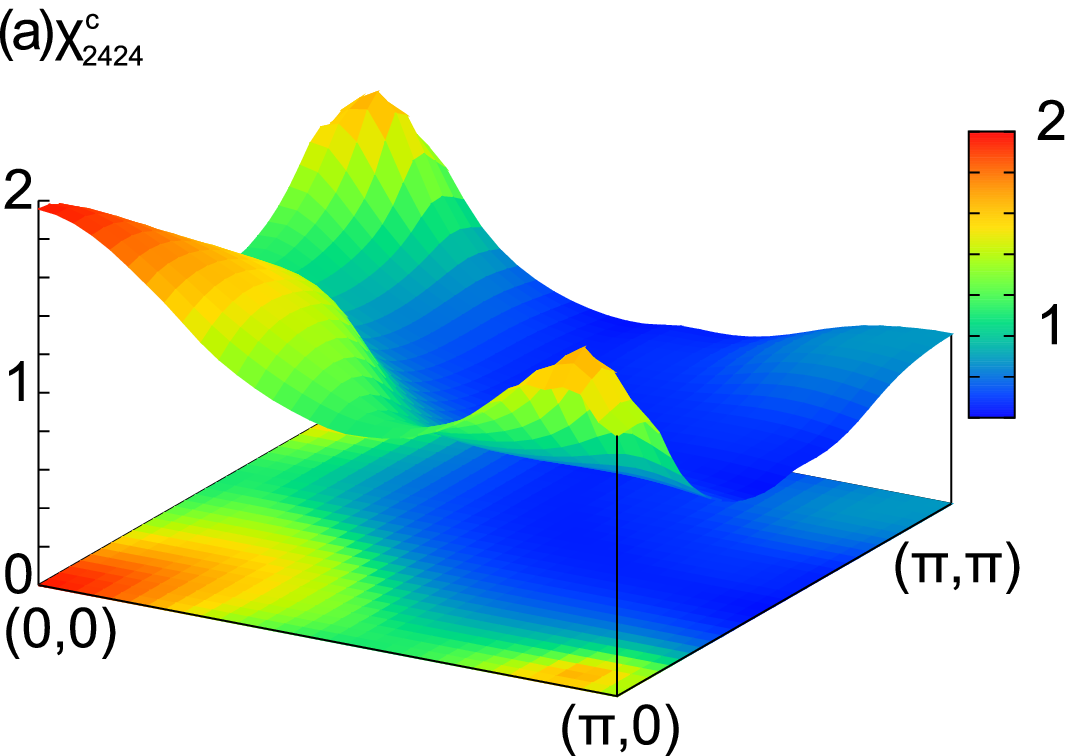}
\includegraphics[width=0.65\linewidth]{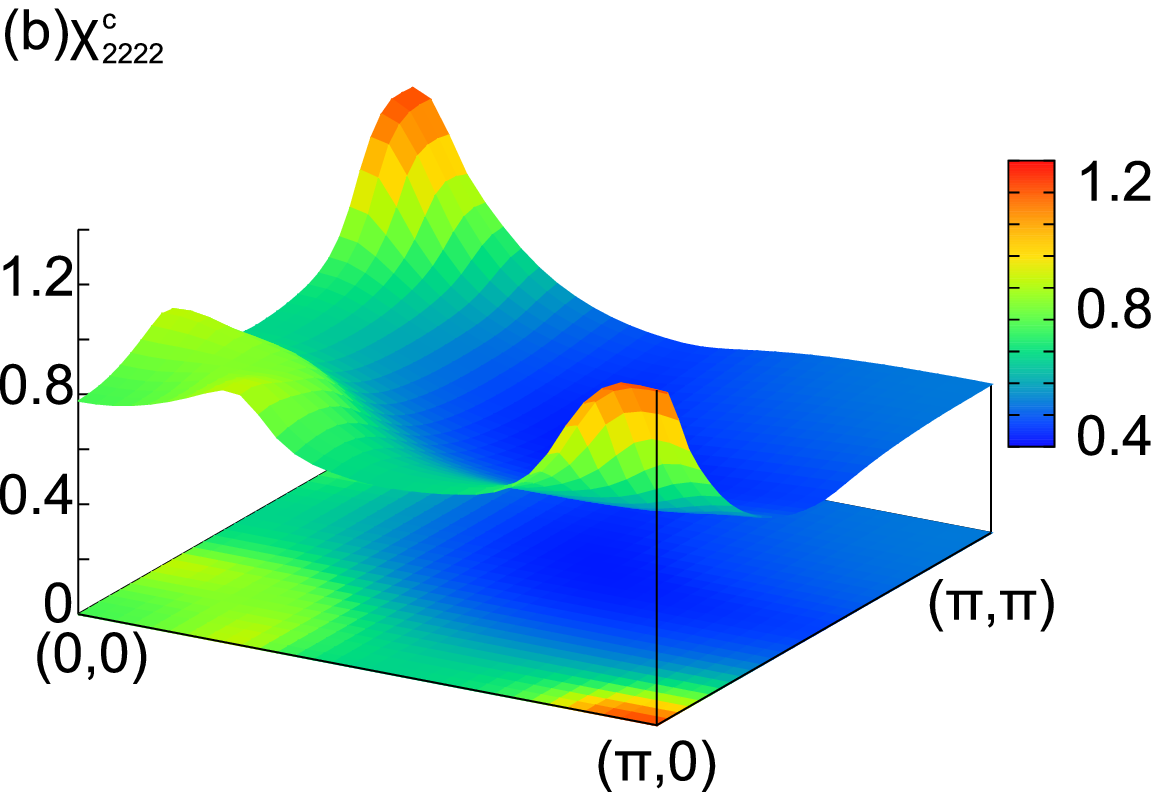}

\caption{
(Color online) 
Obtained (a) $\chi^c_{24,24}( \bm{q} , 0)$ and (b) $\chi^c_{22,22}( \bm{q} , 0)$
for $\alpha=120^{\circ}$, $g(0)=0.21$, $U=1$, and $T=0.02$.
We use 1024 Matsubara frequencies.
}
\label{fig:chi_c(120)}
\end{figure}

Figure \ref{fig:chi_c(120)} shows the obtained $\chi^c_{22,22}( \bm{q} , 0)$ 
and $\chi^c_{24,24}( \bm{q} , 0)$ for $\alpha=120^{\circ}$,
which corresponds to LaFePO.
We put $g(0)=0.21$, $U=1$, and $T=0.02$, which are equivalent to those in Fig. \ref{fig:chi_c}.
Here $\alpha_c=0.98$ is satisfied when As$_4$ tetrahedron is regular.
Compared to Fig. \ref{fig:chi_c} (a), 
magnitude of $\chi_{24,24}^c$ in Fig. \ref{fig:chi_c(120)} (a) is 
largely suppressed due to the reduction in the phonon-mediated interaction 
$g( 0 ,\alpha )$ shown in Fig. \ref{fig:A^2-alpha} (b).

\begin{figure}[!htb]
\includegraphics[width=0.90\linewidth]{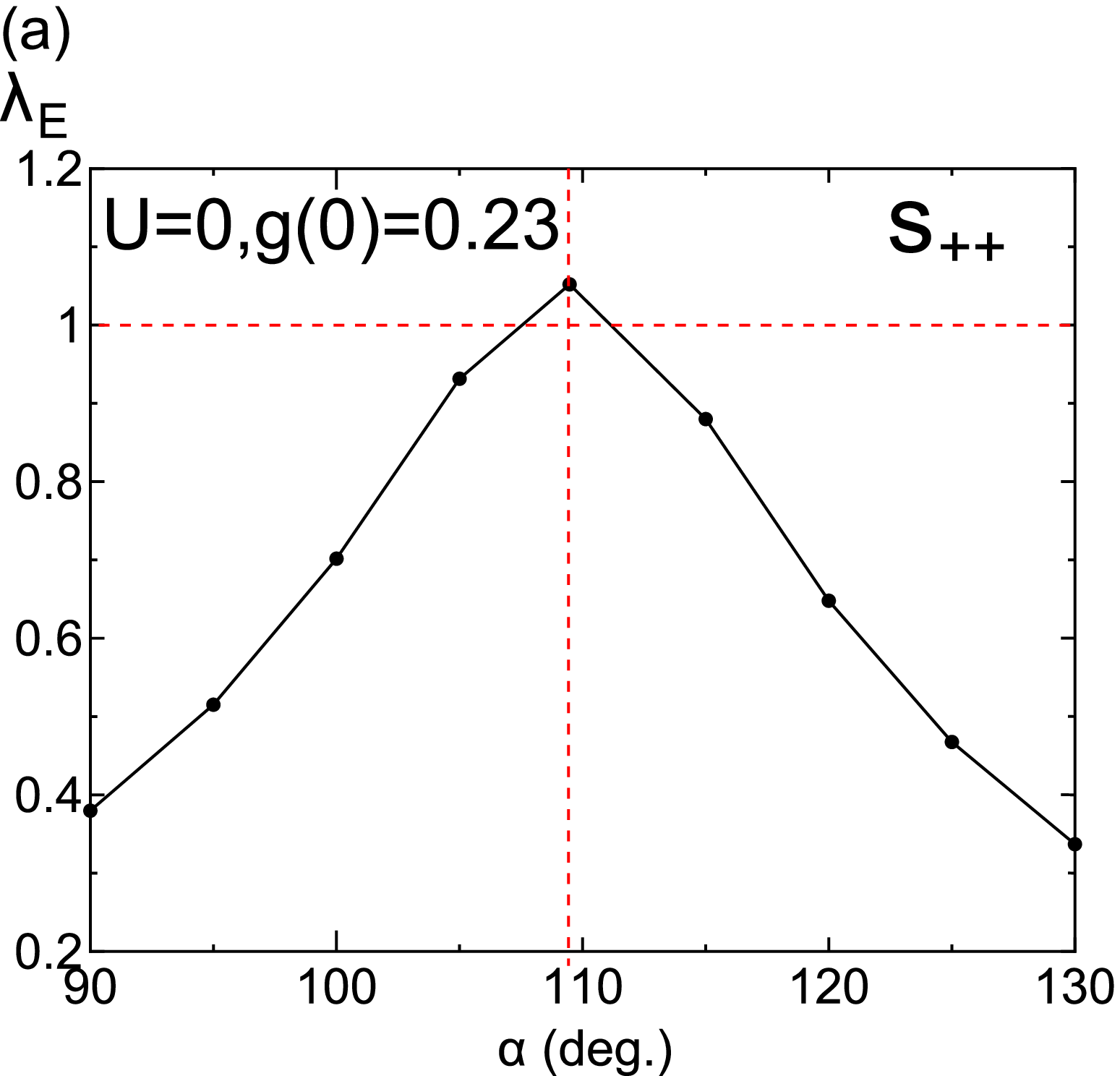}
\includegraphics[width=0.90\linewidth]{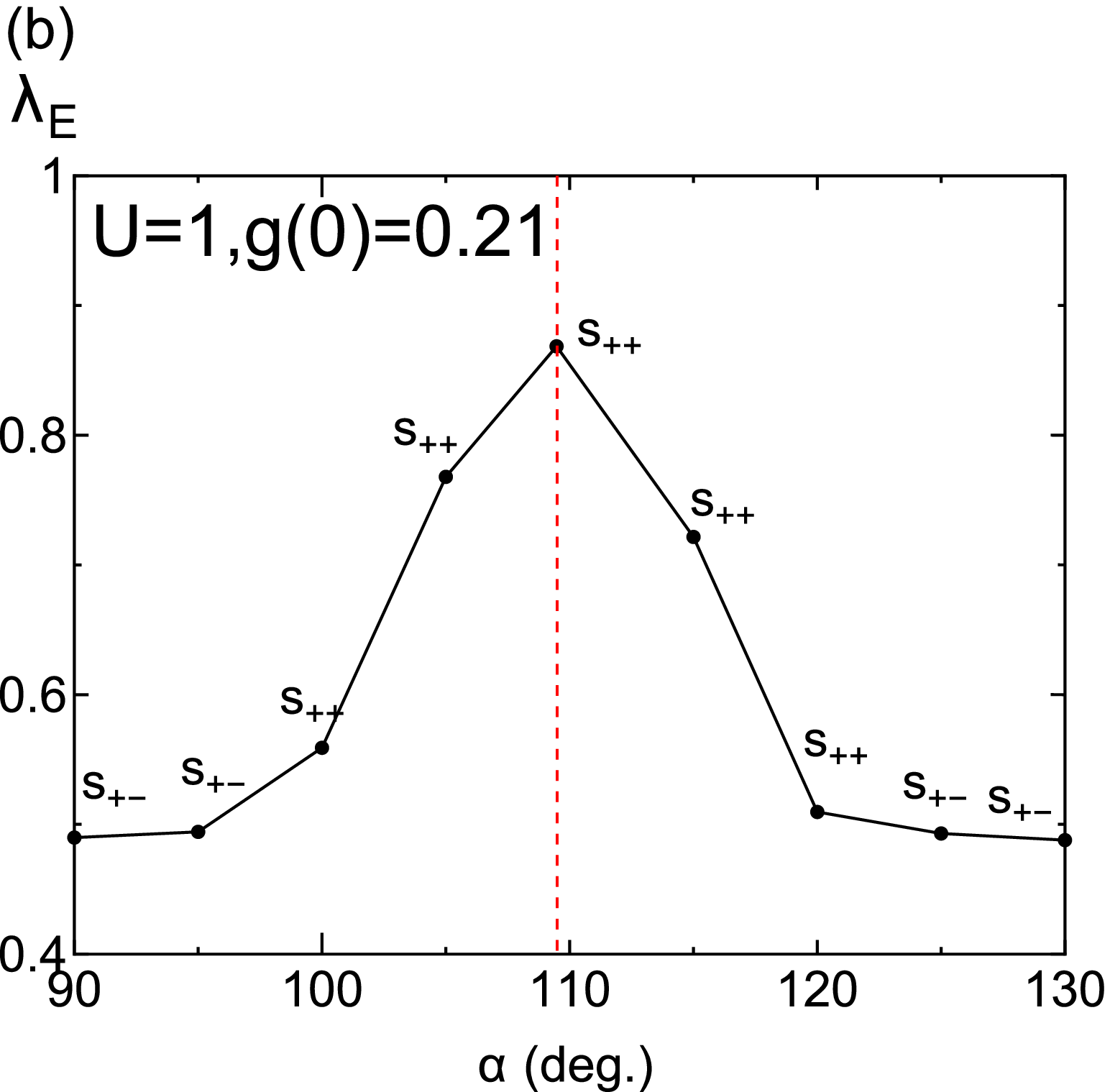}

\caption{
(Color online) 
$\alpha$-dependence of $\lambda_E$ for 
(a)$U=0$ and $g(0)=0.23$, and (b)$U=1$ and $g(0)=0.21$, respectively.
We put $T=0.02$, and $\omega_D = 0.02$.
In (a), $s_{++}$-wave state is always realized.
We use 1024 Matsubara frequencies.
}
\label{fig:lambda-alpha}
\end{figure}

Figure \ref{fig:lambda-alpha} shows the
$\alpha$-dependence of $\lambda_E$ for (a)$U=0$ and $g(0)=0.23$, and (b)$U=1$ and $g(0)=0.21$, respectively.
In both cases, $\alpha_c=0.98$ is satisfied when As$_4$ tetrahedron is regular.
In case (a), the $s_{++}$ state is always realized.
In case (b), the $s_{++}$ state is realized for $100^{\circ} \le \alpha \le 120^{\circ}$,
whereas $s_{\pm}$ is realized for $\alpha \le 95^{\circ}$ or $125^{\circ} \le \alpha$,
because the orbital fluctuations become inferior to spin fluctuations.
In both (a) and (b), $\lambda_E$ for $s_{++}$ state rapidly decreases
when bond angle $\alpha$ deviates from $\alpha_0$.
This result is consistent with the well-known experimental relationship
between bond angle $\alpha$ and $T_{\rm c}$ \cite{Lee,Mizuguchi},
and supports the realization of the orbital-fluctuation-mediated
$s_{++}$-wave state in iron pnictides.

Finally, we note that 
the $e$-ph interaction due to As ion oscillation 
does not take the maximum value at $\a=\a_0$:
For example, $e$-ph interaction by $A_{1g}$-mode,
which is given by the change in the parameter $b$ in Fig. \ref{fig:fig1} (a),
will monotonically increase as $\a$ decreases.
Therefore, the experimental relation between $T_{\rm c}$ and $\a$
strongly suggests that the iron pnictides are not conventional BCS 
superconductors due to charge fluctuations by $A_{1g}$-mode,
but are the orbital-fluctuation-mediated superconductors
due to Fe ion oscillations.

\section{Discussions}

In previous sections,
we have analyzed both $s_{++}$- and  $s_\pm$-wave states
based on the five-orbital HH model.
Here, we discuss both states in more detail,
by making comparison between theoretical results and experimental reports.

\subsection{SC gap in the $Z^2$-orbital hole-pocket
\label{sec:z2}}

Recently, bulk-sensitive ARPES measurements had been performed 
in (Ba,K)Fe$_2$As$_2$ and BaFe$_2$(As,P)$_2$ \cite{Shimojima2}.
The observation have revealed the $Z^2$-orbital hole-pocket 
around $(0,0)$ at $k_z\sim\pi$ in the folded Brillouin zone 
($(\pi,\pi)$ in the unfolded Brillouin zone; see Fig. \ref{fig:gap} (c)),
which was predicted by the first principle LDA study around $k_z=\pi$ 
for larger bond angle $\alpha$ \cite{Kuroki-PRB}.
Moreover, it was found that the magnitude of the SC gap in the 
$Z^2$-orbital hole-pocket is as large as that in other hole-pockets 
composed of $XZ/YZ$- (and $X^2-Y^2$-) orbitals.
However, the SC gap will strongly depend on the orbital nature of 
the FS parts in spin fluctuation mechanism, 
since the $Z^2$-orbital does not participate in the nesting \cite{Shimojima2}.

Therefore, this ``orbital-independent SC gap in 
(Ba,K)Fe$_2$As$_2$ and BaFe$_2$(As,P)$_2$''
is a very crucial test for theories to understand the pairing mechanism.
Here, we shift the $Z^2$-orbital level by $+0.32$eV in the present 
model to reproduce the $Z^2$-orbital hole-pocket 
at $k_z\sim\pi$ in the 3-dimensional model \cite{Kuroki-PRB},
and analyze the orbital dependence of the SC gap in detail.

\begin{figure}[!htb]
\includegraphics[width=0.94\linewidth]{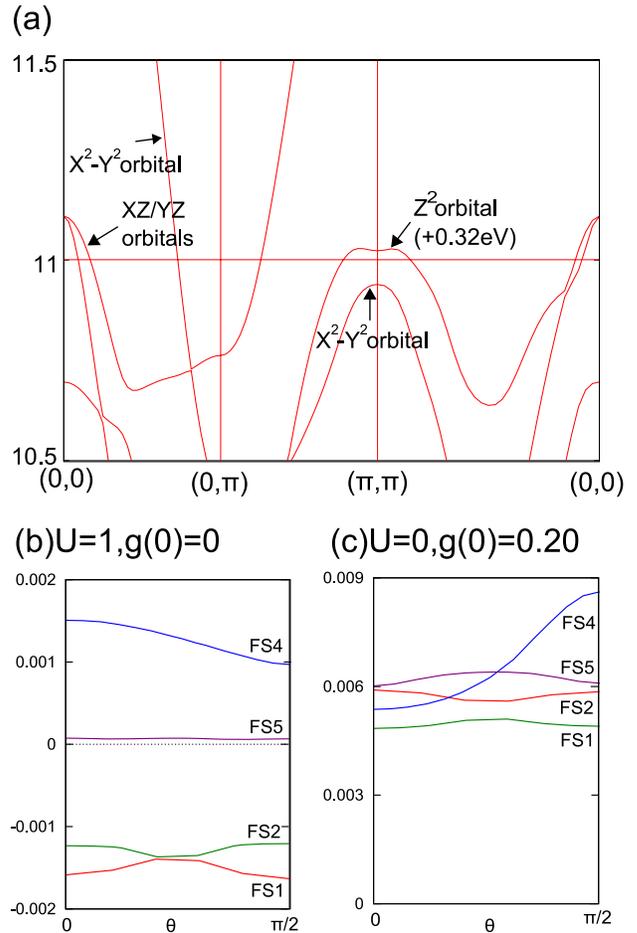}

\caption{
(Color online) 
(a) Band structure (in the unfolded Brillouin zone) given by 
shifting the $Z^2$-orbital level by $+0.32$eV.
The hole-pocket around $(\pi,\pi)$, FS5 in Fig. \ref{fig:gap}(c), 
is composed of $Z^2$-orbital.
(b) $s_\pm$-wave SC gap function for $U=1.0$ and $g(0)=0$.
(c) $s_{++}$-wave SC gap function for $U=0$ and $g(0)=0.19$.
}
\label{fig:z2}
\end{figure}

Figure \ref{fig:z2} (b) shows the 
$s_\pm$-wave SC gap functions obtained for $U=1.0$ and $g(0)=0$.
The obtained parameter is $\lambda_{\rm E}=0.37$. 
(Note that $\lambda_{\rm E}$ for $s_\pm$-wave state 
decreases when $Z^2$-orbital hole pocket appears \cite{Kuroki}.)
As we can see, spin-fluctuation scenario predicts very small SC gap
on the $Z^2$-orbital FS,
since the spin correlation between electrons in $Z^2$-orbital is very small:
In iron pnictides,
spin fluctuations due to the nesting are mainly induced by $XZ/YZ$-orbitals
via intraorbital Coulomb interaction $U$ between opposite spins.
However, spin correlation between different orbitals is much weaker, 
since the Hund's coupling $J$ is much smaller than $U$.
For this reason, the SC gap in the $Z^2$-orbital hole-pocket is very small.
Similar ``orbital dependent SC gap'' is considered to be realized 
in $p$-wave superconductor Sr$_2$RuO$_4$ \cite{Sigrist}.

On the other hand, strong orbital correlation exists for
all $d$-orbitals in the present orbital-fluctuation scenario, 
since the $e$-ph interaction due to Fe-ion oscillation, Eq. (\ref{eqn:Vmat}),
possesses many nonzero interorbital matrix elements.
For this reason, the SC gap in the $Z^2$-orbital hole-pocket can be large.
Figure \ref{fig:z2} (c) shows the 
$s_{++}$-wave SC gap functions obtained for $U=0$ and $g(0)=0.20$
$(\alpha_{\rm c}=0.98)$, induced by orbital fluctuations.
The obtained parameters are $\lambda_{\rm E}=0.99$ and $\alpha_{\rm c}=0.98$.
As expected, the SC gap on the $Z^2$-orbital FS
becomes comparable with that on other FSs.

Therefore, the small orbital-dependence in the SC gap
in (Ba,K)Fe$_2$As$_2$ and BaFe$_2$(As,P)$_2$ \cite{Shimojima2}
supports the present orbital fluctuation scenario.
For a quantitative study of this issue,
large 3-dimensionality of the FSs
in (Ba,K)Fe$_2$As$_2$ and BaFe$_2$(As,P)$_2$ may be important.
This is an important issue for our future investigation.

\subsection{bond angle, pnictogen height}

In Sec. \ref{sec:alpha},
we have explained the experimental relation between
the As-Fe-As bond angle $\alpha$ and $\lambda_{\rm E}$
in iron pnictides by assuming the $s_{++}$-wave SC state
mediated by the orbital fluctuations.
On the other hand, 
Kuroki {\it et al.} \cite{Kuroki-PRB} had studied the same issue
based on the spin fluctuation theory:
In 1111 compounds, the $Z^2$-orbital hole pocket discussed in Sec. \ref{sec:z2}
appears as the pnictogen hight $z=R_{\text{Fe-As}}\cos(\alpha/2)$ decreases. 
Then, $\lambda_{\rm E}$ for $s_\pm$-wave state quickly decreases,
since $Z^2$-orbital does not contributes to the spin fluctuations.
In this scenario,
$\lambda_{\rm E}$ monotonically decreases as $z$ does;
The decrease in $T_{\rm c}$ for $\alpha<\alpha_0$ cannot be explained
without assuming the accidental balance between
pnictogen height effect and another opposite effect.

In the present orbital fluctuation theory,
$\lambda_{\rm E}$ for $s_{++}$-wave state is rather insensitive to 
the appearance of the $Z^2$-orbital hole-pocket
since $Z^2$-orbital also contribute to the formation 
of the orbital fluctuations as discussed in Sec. \ref{sec:z2}.
Therefore, $\lambda_{\rm E}$ or $T_{\rm c}$ will be mainly 
controlled by the bond angle $\alpha$,
which is consistent with the experimental report \cite{Lee}.

\subsection{pressure effect on $T_{\rm c}$}

We also discuss the pressure effect on $T_{\rm c}$ in iron pnictides.
In LaFeAsO$_{1-y}$F$_x$, $T_{\rm c}$ increases from 26 K to 43 K in
overdoped sample ($x=0.14$) by applying $3\sim4$ GPa pressure \cite{Fujiwara}.
In this case,
both $1/T_1T$ and normal-state resistivity $\rho$ around $T_{\rm c}$ are
rather insensitive to the pressure \cite{Fujiwara,Takahashi}.
Similarly, $T_{\rm c}$ increases drastically in FeSe under pressure 
\cite{Mizuguchi,Margadonna}.
On the other hand, $T_{\rm c}$ quickly decreases 
for NdFeAsO$_{1-y}$ and TbFeAsO$_{1-y}$ under pressure,
accompanying the decrease in the temperature dependence of $\rho$
(i.e, the inelastic scattering) \cite{Takeshita}.

To understand the pressure effect on $T_{\rm c}$,
we would have to consider the change in the bandwidth $W_{\rm band}$,
in addition to the bond angle $\alpha$.
In NdFeAsO$_{1-y}$ and TbFeAsO$_{1-y}$,
the reduction in the inelastic scattering in $\rho$ 
under pressure suggests the suppression of spin/orbital fluctuations.
This change is expected to originate from the increase in $W_{\rm band}$, 
which drives the system toward weak-coupling regime.
Then, reduction in spin/orbital fluctuations under pressure
should make $T_{\rm c}$ lower.

Next, we discuss the possible origin of the enhancement in $T_{\rm c}$ 
under pressure.
In FeSe, $T_{\rm c}$ increases under pressure $\sim8$GPa,
whereas $\alpha_0-\alpha(>0)$ slightly increases.
Here, we emphasize that the change in the Fe-As bond length $R_{\text{Fe-As}}$
would be the key parameter:
Under pressure, the effective interaction due to $e$-ph coupling is given by 
$g(0)=(R_{\text{Fe-As}}^0/R_{\text{Fe-As}})^8g^0(0)$,
where the suffix $0$ represents the quantity at ambient pressure.
According to Ref. \cite{Margadonna},
$R_{\text{Fe-As}}^0/R_{\text{Fe-As}}=2.38{\buildrel _{\circ} \over {\mathrm{A}}}
/2.30{\buildrel _{\circ} \over {\mathrm{A}}}$ at 8GPa,
and thus $g(0)\sim(4/3)g^0(0)$.
This prominent enhancement in $g(0)$ under pressure might be 
the origin of strong increase of $T_{\rm c}$ in FeSe under pressure.

\subsection{iron isotope effect}

In Sec. \ref{sec:Eliash},
we have discussed the iron isotope effect
based on the orbital fluctuation scenario.
Experimentally, Liu {\it et al.} reported that
the iron isotope coefficient $\beta =-\d \ln{T_c} / \d \ln M$ is
$\sim0.35$ for SmFeAsO$_{0.85}$F$_{0.15}$ and Ba$_{0.6}$K$_{0.4}$Fe$_2$As$_2$
 \cite{Liu}.
However, Shirage {\it et al.} had recently reported the 
negative (or zero) iron isotope effect for the same compounds;
$\beta\sim-0.18$ for (Ba,K)Fe$_2$As$_2$ \cite{Shirage1} 
and $\beta\sim-0.02$ for SmFeAsO$_{y}$ \cite{Shirage2}.
The reason for the discrepancy is yet unclear.

We have shown in Sec. \ref{sec:Eliash} that the coefficient $\beta$
changes from positive to negative as the Coulomb interaction increases.
The first principle calculations had estimated that 
the ratio $U/W_{\rm band}$ for 122 compounds is larger than that 
for 1111 compounds \cite{Arita}.
Then, negative (zero) isotope effect reported for 
(Ba,K)Fe$_2$As$_2$ (SmFeAsO$_{y}$) does not contradict with 
the orbital fluctuation scenario.

In the spin fluctuation scenario, $\beta$ becomes positive (negative)
when the intra-pocket phonon-mediated attractive interaction $g({\bm 0};\w_l)$
is superior (inferior) to the inter-pocket one $g({\bm Q};\w_l)$
 \cite{Yanagisawa,Bang}.
Therefore, sign change in $\beta$ can be explained 
if the nature of $e$-ph interaction largely
depends on the compounds.

\section{Summary}
In the present paper,
we studied the five-orbital HH model for iron pnictides, 
and found that $s_{++}$-wave SC state is induced by orbital fluctuations
in the presence of small $e$-ph interaction ($\lambda\lesssim0.15$).
Strong orbital fluctuations are induced by multiple scattering processes 
due to the $e$-ph interaction, involving all five $d$-orbitals on the FSs.
We stress that the second-order process alone, 
which is usually studied in conventional BCS analysis \cite{lambda-LDA},
can neither induce large orbital fluctuations nor 
high-$T_{\rm c}$ $s_{++}$-wave SC state.
Roughly speaking, $T_{\rm c}$ in the orbital fluctuation theory would be
given as $T_{\rm c}\sim\w_{\rm D}\exp(-1/\lambda^*)$, where 
$\lambda^*\sim\lambda(1-\alpha_{\rm c})^{-1}$ is the enhanced coupling constant,
and thus it is much larger than $T_{\rm c}^{\rm BCS}\sim\w_{\rm D}\exp(-1/\lambda)$.

The virtue of this theory is that we can also explain the following 
issues which remain unresolved within the spin fluctuation theory:
(i) empirical relationship between $T_{\rm c}$ 
and the As-Fe-As bond angle (Lee plot),
(ii) negative iron isotope effect in (Ba,K)Fe2As2, and
(iii) orbital-independent SC gap in (Ba,K)Fe$_2$As$_2$ and BaFe$_2$(As,P)$_2$ 
observed by bulk-sensitive ARPES measurement \cite{Shimojima2}.
Recently, theoretically predicted orbital fluctuations 
in Figs. \ref{fig:chi_c} (a) and (b)
had been confirmed by the softening of the elastic constants
$C_{44}$ and $C_{\rm E}$ \cite{Yoshizawa}.
These obtained results support the idea of the $s_{++}$-wave state 
mediated by orbital fluctuations in iron pnictides,
next to the orbital-ordered state in mother compounds.

Finally, we list several significant future issues.
The $e$-ph interaction due to corrective oscillations 
(e.g., half-breathing mode) might be important to increase 
the $s_{++}$-wave $T_{\rm c}$.
To make quantitative comparison between $s_{++}$-wave and $s_\pm$-wave states,
study of the self-energy and vertex corrections
for $\chi^{c(s)}$ and the Eliashberg gap equation is highly desired.
The FLEX approximation would be useful for this purpose
\cite{Onari-future}.


\acknowledgements
We are grateful to 
D.S. Hirashima, M. Sato, Y. Kobayashi, Y. Matsuda, T. Shibauchi,
M. Takigawa, S. Shin, T. Shimojima, Y. ${\bar {\rm O}}$no,
F.C. Zhang, and M. Yoshizawa for useful comments and discussions.
This study has been supported by Grants-in-Aid for Scientific 
Research from MEXT of Japan, and by JST, TRIP.
Numerical calculations were performed using the facilities of 
the supercomputer center, Institute for Molecular Science.

\appendix

\section{Phase Diagram and Orbital Fluctuations for $J/U>1/6$}

\begin{figure}[!htb]
\includegraphics[width=0.80\linewidth]{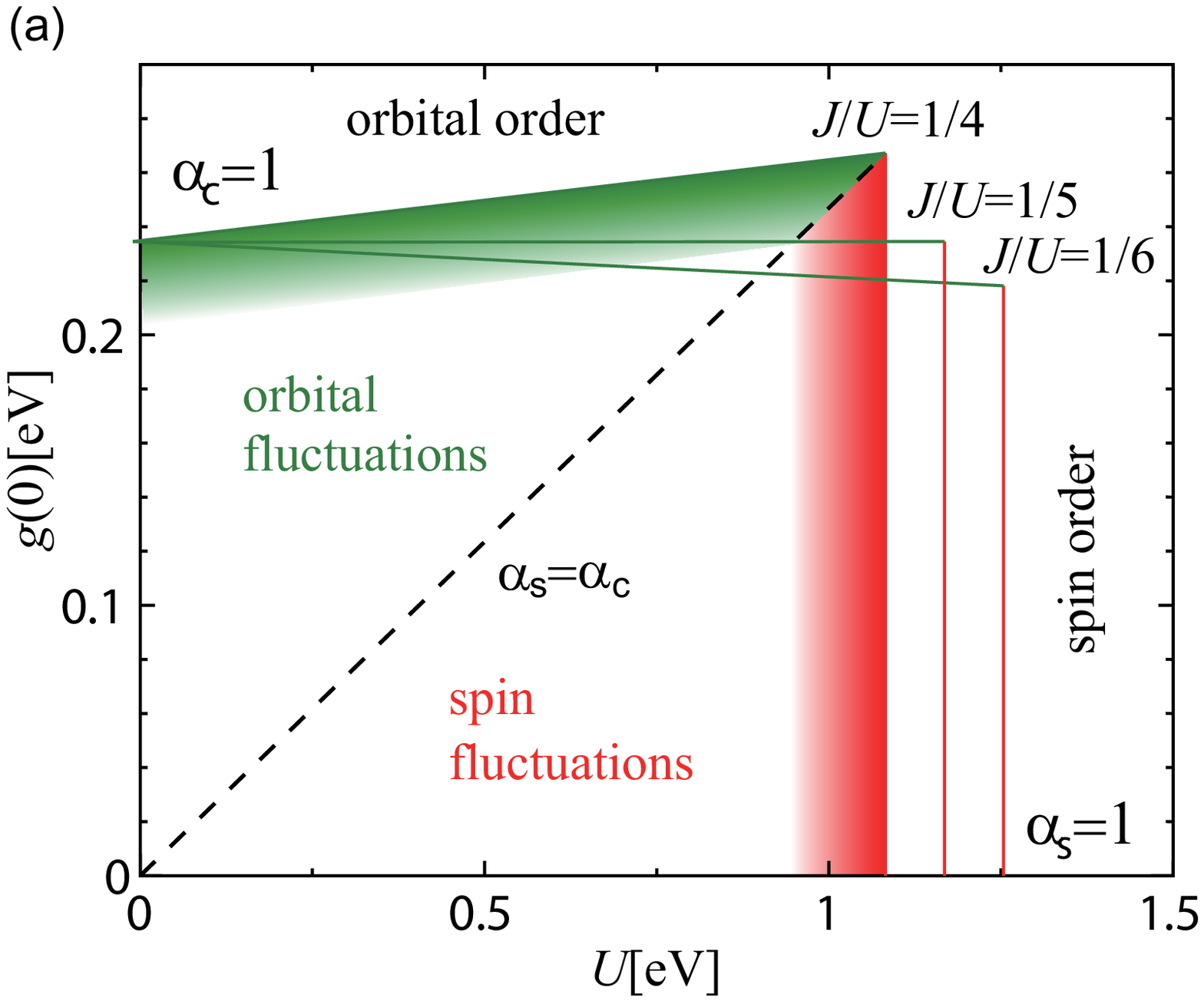}
\includegraphics[width=0.94\linewidth]{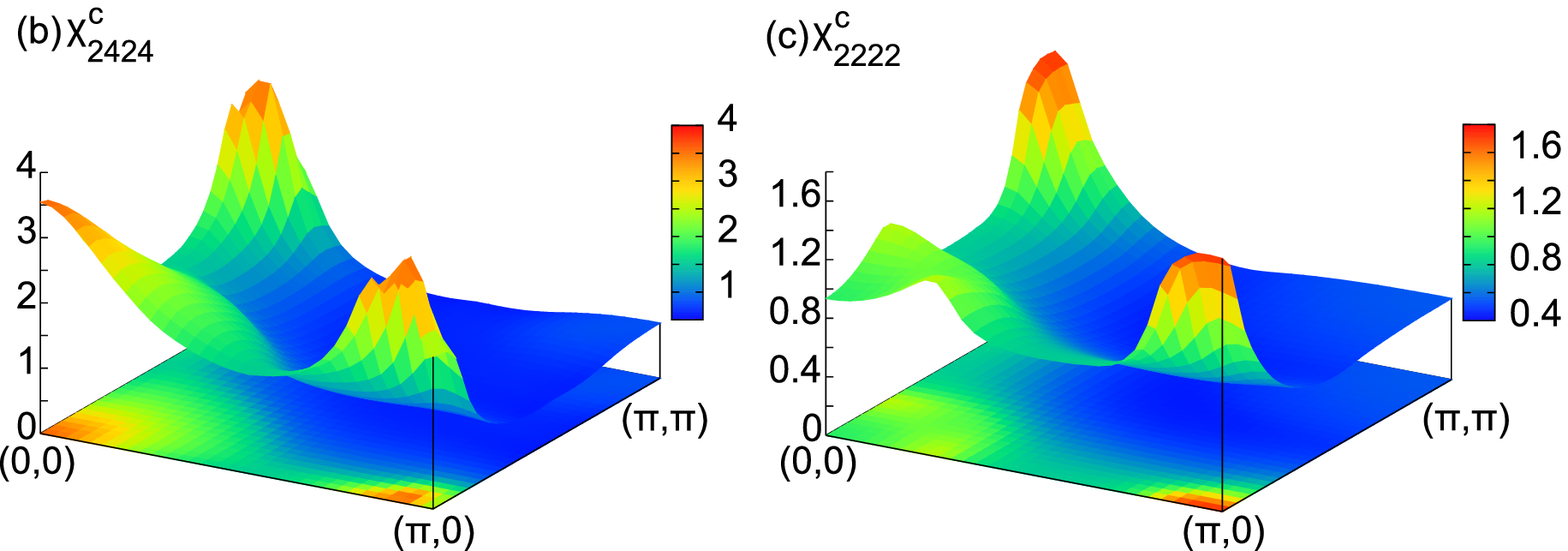}
\caption{
(Color online) 
(a) Obtained $U$-$g(0)$ phase diagram for $J/U=1/4$, $1/5$ and $1/6$. 
(b) Obtained $\chi_{24,24}^c(\bm{q},0)$ for $J/U=1/6$ and $n=6.1$.
(c) Obtained $\chi_{22,22}^c(\bm{q},0)$ for $J/U=1/6$ and $n=6.1$.
}
\label{fig:Ap1}
\end{figure}

In the present paper, we studied the multiorbital HH model 
for general parameters under the constraint $J/U=1/6$.
In fact, based on the first-principle calculation,
Miyake {\it et al.} had derived the averaged $J/U$ as 0.4eV/2.5eV=1/6.3 
for LaFeAsO and 0.45eV/3eV=1/6.7 for BaFe$_2$As$_2$, respectively
\cite{Arita}.
Moreover, very small value of $J/U<0.1$ is required to reproduce
the small magnetic moment in the SDW state
within the mean-field approximation \cite{MF}.
However, $J/U$ is expected to be larger for usual iron-compounds,
and there is no consensus on the value of $J/U$ in iron pnictides 
up to now.
In case of $J/U=1/6$, Coulomb interaction 
enhances the orbital fluctuations when $g(0)$ is fixed,
as shown in Fig. \ref{fig:U-g0}.
However, it is highly desired to 
study the orbital fluctuation for general value of $J/U$.
In this appendix, we discuss this issue using the RPA, 
and show that characteristic nature of orbital fluctuations 
does not influenced by $J/U$.

Figure \ref{fig:Ap1} (a) represents the $U$-$g(0)$ phase diagram
given by the mean-field approximation for $J/U=1/4$, $1/5$ and $1/6$.
For $J/U=1/4$, the Coulomb interaction 
reduces the charge Stoner factor $\a_c$ when $g(0)$ is fixed,
indicating the suppression of orbital fluctuations.
The value of $g_{\rm cr}(0)$ for $\a_c=1$ is 0.26 when $U=1.0$;
the corresponding dimensionless coupling $\lambda\equiv gN(0)$
is only 0.18.
For $J/U=1/5$, $\a_c$ is almost independent of $U$.
Therefore, the value of $g_{\rm cr}(0)$ increases with $J/U$:
The reason is that $U$ ($U'=U-2J$) reduces (enhances) 
orbital fluctuations.
As for the spin correlation,
the value of $U_{\rm cr}$ for $\a_s=1$ decreases with $J/U$,
indicating the enhancement of spin fluctuations.

Figure \ref{fig:Ap1} (b) and (c) show the obtained 
$\chi^c_{24,24}( \bm{q},0)$ and $\chi^c_{22,22}( \bm{q},0)$
for $J/U=1/4$, $U=1$, and $a_c=0.98$ ($g(0)\approx0.26$) at $T=0.02$.
Comparing with Figs. \ref{fig:chi_c} (a) and (b),
it is found that the orbital susceptibilities are almost
independent of $J/U\le1/4$ for a fixed $\a_c$.
Thus, the present orbital fluctuation scenario for iron pnictides
would be plausible for $J/U\le1/4$.



\begin{thebibliography}{99}

\bibitem{Kambara}
Y. Kamihara, T. Watanabe, M. Hirano, and H. Hosono,
J. Am. Chem. Soc. {\bf 130}, 3296 (2008). 

\bibitem{Kuroki}
K. Kuroki, S. Onari, R. Arita, H. Usui, Y. Tanaka, H. Kontani and H. Aoki,
Phys. Rev. Lett. {\bf 101}, 087004 (2008).

\bibitem{Mazin}
I. I. Mazin, D. J. Singh, M. D. Johannes, and M. H. Du,
Phys. Rev. Lett. {\bf 101}, 057003 (2008).

\bibitem{Moriya}
T. Moriya, Y. Takahashi, and K. Ueda, J. Phys. Soc. Jpn.
{\bf 59}, 2905 (1990).
K. Ueda, T. Moriya and Y. Takahashi, J. Phys. Chem.
Solids. {\bf 53}, 1515 (1992).

\bibitem{Pines}
P. Monthoux and D. Pines, Phys. Rev. B {\bf 47}, 6069 (1993).

\bibitem{Bickers}
N. E. Bickers and S. R. White, Phys. Rev. B {\bf 43}, 8044 (1991).

\bibitem{Schmalian}
J. Schmalian, Phys. Rev. Lett. {\bf 81}, 4232 (1998).

\bibitem{Kino}
H. Kino and H. Kontani, J. Phys. Soc. Jpn. {\bf 67}, 3691 (1998).

\bibitem{Kondo}
H. Kondo and T. Moriya, J. Phys. Soc. Jpn. {\bf 67}, 3695 (1998).

\bibitem{Takimoto-Ce115}
T. Takimoto, T. Hotta, and K. Ueda,
Phys. Rev. B {\bf 69}, 104504 (2004).

\bibitem{Onari-impurity}
S. Onari, and H. Kontani,
Phys. Rev. Lett. {\bf 103}, (2009) 177001.

\bibitem{Sato-imp}
A. Kawabata, S. C. Lee, T. Moyoshi, Y. Kobayashi and M. Sato,
J. Phys. Soc. Jpn. {\bf 77}, 103704 (2008);
M. Sato, Y. Kobayashi, S. C. Lee, H. Takahashi, E. Satomi and Y. Miura, J. Phys. Soc. Jpn. {\bf 79}, 014710 (2009);
S.C. Lee, E. Satomi, Y. Kobayashi, and M. Sato, J. Phys. Soc. Jpn. {\bf 79}, 023702 (2010).

\bibitem{irradiation}
C. Tarantini, M. Putti, A. Gurevich, Y. Shen, R. K. Singh, J. M. Rowell, N. Newman, D. C. Larbalestier,
P. Cheng, Y. Jia, and H.-H. Wen, Phys. Rev. Lett. {\bf 104}, 087002 (2010).

\bibitem{Nakajima}
Y. Nakajima, T. Taen, Y. Tsuchiya, T. Tamegai, H. Kitamura, and T. Murakami,
arXiv:1009.2848.

\bibitem{FCZhang}
Y. Li, J. Tong, Q. Tao, C. Feng, G. Cao, W. Chen, F.C. Zhang, and Z.A. Xu,
New J. Phys. {\bf 12} (2010) 083008.

\bibitem{iikubo-sato}
S. Iikubo, M. Ito, A. Kobayashi, M. Sato and K. Kakurai, J. Phys. Soc. Jpn. {\bf 74}, 275 (2005).

\bibitem{ito-sato}
M. Ito, H. Harashina, Y. Yasui, M. Kanada, S. Iikubo, M. Sato, A. Kobayashi and K. Kakurai,
J. Phys. Soc. Jpn. {\bf 71}, 265 (2002).

\bibitem{keimer-highTc}
H. F. Fong, P. Bourges, Y. Sidis, L. P. Regnault, A. Ivanov, G. D. Gu, N. Koshizuka and B. Keimer, Nature {\bf 398}, 588 (1999).

\bibitem{res-115}
C. Stock, C. Broholm, J. Hudis, H. J. Kang, and C. Petrovic, Phys. Rev. Lett. {\bf 100}, 087001 (2008).

\bibitem{christianson}
A. D. Christianson, E. A. Goremychkin, R. Osborn, S. Rosenkranz, M. D. Lumsden, C. D. Malliakas, I. S. Todorov, H. Claus,
D. Y. Chung, M. G. Kanatzidis, R. I. Bewley and T. Guidi, Nature {\bf 456}, 930 (2008).

\bibitem{qiu}
Y. Qiu, W. Bao, Y. Zhao, C. Broholm, V. Stanev, Z. Tesanovic, Y. C. Gasparovic, S. Chang,
J. Hu, B. Qian, M. Fang and Z. Mao, Phys. Rev. Lett. {\bf 103}, 067008 (2009).

\bibitem{keimer}
D. S. Inosov, J. T. Park, P. Bourges, D. L. Sun, Y. Sidis, A. Schneidewind, K. Hradil, D. Haug,
C. T. Lin, B. Keimer and V. Hinkov, Nature Physics {\bf 6} 178 (2010).

\bibitem{Onari-resonance}
S. Onari, H. Kontani and M. Sato,
Phys. Rev. B 81, 060504(R), (2010).

\bibitem{Ishida}
Y. Nakai, T. Iye, S. Kitagawa, K. Ishida, H. Ikeda, S. Kasahara, 
H. Shishido, T. Shibauchi, Y. Matsuda, and T. Terashima,
Phys. Rev. Lett. {\bf 105}, 107003 (2010).

\bibitem{Fujiwara}
T. Nakano, N. Fujiwara, K. Tatsumi, H. Okada, H. Takahashi, Y. Kamihara, M. Hirano, and H. Hosono,
Phys. Rev. B {\bf 81}, 100510(R) (2010).

\bibitem{Kontani}
H. Kontani and S. Onari, 
Phys. Rev. Lett. {\bf 104}, 157001 (2010).

\bibitem{Yanagi}
Y. Yanagi and Y. Yamakawa, and Y. ${\bar {\rm O}}$no,
Phys. Rev. B {\bf 81}, 054518 (2010).

\bibitem{lambda-LDA}
L. Boeri, O. V. Dolgov, A. A. Golubov,
Phys. Rev. Lett. {\bf 101}, 026403 (2008).

\bibitem{softening}
R. M. Fernandes, L. H. VanBebber, S. Bhattacharya, P. Chandra, V. Keppens, D. Mandrus, M. A. McGuire,
B. C. Sales, A. S. Sefat, J. Schmalian, arXiv:0911.3084.

\bibitem{Yoshizawa}
M. Yoshizawa, R. Kamiya, R. Onodera, Y. Nakanishi,
K. Kihou, H. Eisaki, and C. H. Lee, arXiv:1008.1479.

\bibitem{Raman}
M. Rahlenbeck, G. L. Sun, D. L. Sun, C. T. Lin, B. Keimer, C. Ulrich,
Phys. Rev. B {\bf 80}, 064509 (2009).

\bibitem{Dresden}
A. A. Kordyuk, V. B. Zabolotnyy, D. V. Evtushinsky, T. K. Kim, 
I. V. Morozov, M. L. Kulic, R. Follath, G. Behr, B. Buechner, 
and S. V. Borisenko,
arXiv:1002.3149.

\bibitem{optical}
T. Dong, Z. G. Chen, R. H. Yuan, B. F. Hu, B. Cheng, and N. L. Wang,
arXiv:1005.0780.

\bibitem{Lee}
C.-H. Lee, A. Iyo, H. Eisaki, H. Kito, M. T. Fernandez-Diaz, T.
Ito, K. Kihou, H. Matsuhata, M. Braden, and K. Yamada,
J. Phys. Soc. Jpn. {\bf 77}, 083704 (2008).

\bibitem{Shirage1}
P. M. Shirage, K. Kihou, K. Miyazawa, C.-H. Lee, H. Kito, H. Eisaki,
T. Yanagisawa, Y. Tanaka and A. Iyo,
Phys. Rev. Lett. {\bf 103}, 257003 (2009).

\bibitem{Shimojima2}
T. Shimojima, 
private communication.

\bibitem{Shimojima}
T. Shimojima, K. Ishizaka, Y. Ishida, N. Katayama, K. Ohgushi, T. Kiss, M. Okawa, T. Togashi,
X.-Y. Wang, C.-T. Chen, S. Watanabe, R. Kadota, T. Oguchi, A. Chainani, and S. Shin, Phys. Rev. Lett. {\bf 104}, 057002 (2010).

\bibitem{detwinned}
J.-H. Chu, J. G. Analytis, K. D. Greve, P. L. McMahon, Z. Islam, 
Y. Yamamoto, I. R. Fisher, 
arXiv:1002.3364;
A. Dusza, A. Lucarelli, F. Pfuner, J.-H. Chu, I. R. Fisher, and L. Degiorgi,
arXiv:1007.2543.

\bibitem{Takimoto}
T. Takimoto, T. Hotta, T. Maehira and K. Ueda,
J. Phys. Condens. Matter {\bf 14}, L369 (2002).

\bibitem{Onari-future}
S. Onari and H. Kontani, arXiv:1009.3882.

\bibitem{Matsuda}
K. Hashimoto, M. Yamashita, S. Kasahara, Y. Senshu, N. Nakata, S. Tonegawa, K. Ikada, A. Serafin, A. Carrington,
T. Terashima, H. Ikeda, T. Shibauchi and Y. Matsuda, Phys. Rev. B {\bf 81}, 220501(R) (2010).

\bibitem{Morel}
P. Morel and P. W. Anderson, 
Phys. Rev. {\bf 125}, 1263 (1962).

\bibitem{Allen}
P. B. Allen and B. Mitrovi\'{c},
Solid State Physics {\bf 37}, 1 (1982).

\bibitem{Kontani-imp}
H. Kontani and M. Ohno, Phys. Rev. B {\bf 74}, 014406 (2006).

\bibitem{Mizuguchi}
Y. Mizuguchi, Y. Hara, K. Deguchi, S. Tsuda,
T. Yamaguchi, K. Takeda, H. Kotegawa, H. Tou and
Y. Takano,
Supercond. Sci. Technol. {\bf 23}, 054013 (2010).

\bibitem{Kuroki-PRB}
K. Kuroki, H. Usui, S. Onari, R. Arita, H. Aoki,
Phys. Rev. B {\bf 79}, 224511 (2009). 

\bibitem{Sigrist}
D.F. Agterberg, T.M. Rice, and M. Sigrist,
Phys. Rev. Lett. {\bf 78}, 3374 (1997).

\bibitem{Takahashi}
T. Takahashi, K. Igawa, K. Arii, Y. Kamihara, M. Hirano and H. Hosono, Nature {\bf 453}, 376 (2008).

\bibitem{Margadonna}
S. Margadonna, Y. Takabayashi, Y. Ohishi, Y. Mizuguchi,
Y. Takano, T. Kagayama, T. Nakagawa, M. Takata
and K. Prassides,
Phys. Rev. B {\bf 80}, 064506 (2009).

\bibitem{Takeshita}
N. Takeshita, A. Iyo, H. Eisaki, H. Kito, and T. Ito, J. Phys. Soc. Jpn. {\bf 77} 075003 (2008).

\bibitem{Liu}
R. H. Liu, T. Wu, G. Wu, H. Chen, X. F. Wang, Y. L. Xie, J. J. Ying, Y. J. Yan, Q. J. Li, B. C. Shi, W. S. Chu,
Z. Y. Wu and X. H. Chen,
Nature {\bf 459}, 64 (2009).

\bibitem{Shirage2}
P. M. Shirage, K. Miyazawa, K. Kihou, H. Kito, Y. Yoshida, Y. Tanaka,
H. Eisaki and A. Iyo,
Phys. Rev. Lett. {\bf 105}, 037004 (2010).

\bibitem{Arita}
T. Miyake, K. Nakamura, R. Arita and M. Imada,
J. Phys. Soc. Jpn. {\bf 79} 044705 (2010).

\bibitem{Yanagisawa}
T. Yanagisawa, K. Odagiri, I. Hase, K. Yamaji, P. M. Shirage, Y. Tanaka, A. Iyo and H. Eisaki,
J. Phys. Soc. Jpn. {\bf 78}, 094718 (2009).

\bibitem{Bang}
Y. Bang,
Phys. Rev. B {\bf 79}, 092503 (2009).

\bibitem{MF}
E. Bascones, M.J. Calderon, and B. Valenzuela,
Phys. Rev. Lett. {\bf 104}, 227201 (2010).

\end{thebibliography}
\end{document}